\documentclass[twocolumn,superscriptaddress,amsfont,amssymb,amsmath, showpacs,balancelastpage, nofootinbib]{revtex4-1}
\usepackage{graphicx,longtable,mathrsfs,color,array}
\usepackage[hidelinks]{hyperref}
\usepackage[usenames,dvipsnames]{xcolor}
\usepackage{amssymb,amsmath,mathtools,mathrsfs}
\usepackage{epsfig,subfigure,placeins,float}
\usepackage{booktabs,longtable,ctable,multirow}
\usepackage{exscale,relsize}
\usepackage[normalem]{ulem}
\usepackage{enumerate}
\usepackage{times, mathptmx}
\usepackage{colortbl} 
\usepackage[dvipsnames]{xcolor} 
\usepackage{pifont} 
\usepackage{amsmath} 
\usepackage{adjustbox}

\newcommand{\be}{\begin{equation}}
\newcommand{\ee}{\end{equation}}
\def\Mp{M_{\rm Pl}}

\newcommand{\mpl}{M_{\rm Pl}}

\newcommand\alphaB{\alpha_{\text{B}}}
\newcommand\alphaM{\alpha_{\text{M}}}

\newcommand\alphaT{\alpha_{\text{T}}}
\newcommand\alphaH{\alpha_{\text{H}}}
\newcommand\alphaV{\alpha_{\text{V}}}
\def\d{\delta}

\newcommand{\cL}{\mathcal{L}}

\newcommand{\eqn}[1]{eq.~(\ref{#1})}

\newcommand{\appref}[1]{App.~\ref{#1}}
\newcommand{\secref}[1]{Sec.~\ref{#1}}
\newcommand{\half}{\frac{1}{2}}

  \newcommand{\figref}[1]{Fig.~\ref{#1}}

\newcommand{\bun}{\beta_1}
\newcommand{\bdeux}{\beta_2}
\newcommand{\btrois}{\beta_3}

\newcommand{\nuone}{\nu_1}
\newcommand{\nutwo}{\nu_2}

\newcommand{\nufour}{\nu_3}
\newcommand{\nufive}{\nu_4}
\newcommand{\nusix}{\nu_5}
\newcommand{\epsilonMatt}{\epsilon}

\allowdisplaybreaks

\begin{document}

\title{{Vainshtein regime in Scalar-Tensor gravity: constraints on DHOST theories}}
\author{Marco Crisostomi}
\affiliation{Institut de physique th\' eorique, Universit\'e  Paris Saclay 
CEA, CNRS, 91191 Gif-sur-Yvette, France}
\affiliation{AIM, CEA, CNRS, Univ.\ Paris-Saclay, Univ.\ Paris Diderot,
Sorbonne Paris Cit\'e, F-91191 Gif-sur-Yvette, France}
\affiliation{Laboratoire de Physique Th\'eorique, CNRS, Univ.\ Paris-Sud, 
Universit\'e Paris-Saclay, 91405 Orsay, France}
\author{Matthew Lewandowski}
\affiliation{Institut de physique th\' eorique, Universit\'e  Paris Saclay 
CEA, CNRS, 91191 Gif-sur-Yvette, France}
\author{Filippo Vernizzi}
\affiliation{Institut de physique th\' eorique, Universit\'e  Paris Saclay 
CEA, CNRS, 91191 Gif-sur-Yvette, France}
        \date{\today}

\begin{abstract}

We study the screening mechanism in the most general scalar-tensor theories that leave gravitational waves unaffected and are thus compatible with recent LIGO/Virgo observations. 
Using the effective field theory of dark energy approach, we consider the general action for perturbations beyond linear order, focussing on the quasi-static limit. When restricting to 
the subclass of theories that satisfy the gravitational wave constraints,
the fully nonlinear effective {Lagrangian contains only three independent parameters. One of these, $\beta_1$, is uniquely present in degenerate higher-order theories.  
We compute the two gravitational potentials for a spherically symmetric matter source and we find that for  $\beta_1 \ge 0$ they decrease as the inverse of the distance, as in standard gravity,  while the case $\beta_1 < 0$ is ruled out. For $\beta_1 > 0$, the two potentials differ and their 
gravitational 
constants 
are not the same on the inside and outside of the body.}
Generically, the  bound on anomalous light {bending} in the Solar-System constrains $\beta_1 \lesssim 10^{-5}$. Standard gravity can be recovered outside the body by tuning the parameters of the model, in which case $\beta_1 \lesssim 10^{-2}$ from the Hulse-Taylor pulsar.
{Theories conformally related to General Relativity admit $0 \le \beta_1 \lesssim 10^{-6}$, at least for a specific choice of conformal couplings.}

\end{abstract}

\maketitle

\section{Introduction}

Scalar-tensor theories are currently used to extend gravity beyond General Relativity on cosmological scales. The higher-derivate terms that characterize Horndeski  \cite{Horndeski:1974wa,Deffayet:2011gz} and beyond-Horndeski theories \cite{Zumalacarregui:2013pma,Gleyzes:2014dya,Langlois:2015cwa,Crisostomi:2016czh,BenAchour:2016fzp}  (see also \cite{Langlois:2018dxi,Kobayashi:2019hrl} for recent reviews), while possibly providing an origin for the observed accelerated expansion of the Universe, are crucial to suppress modifications of gravity on  Solar-System, where very stringent tests apply \cite{Will:2014xja}.   

 {Indeed, close to matter sources, the nonlinearities of the scalar field suppress its effects through the so-called Vainshtein screening mechanism  \cite{Vainshtein:1972sx,Babichev:2013usa}, allowing General Relativity to be recovered.  Screening is essential} for the theories under consideration to be observationally viable.

However, recent observations of gravitational waves have put severe constraints on higher-derivative operators. The  simultaneous observation of gravitational waves
and gamma ray bursts from the GW170817 event has  constrained very precisely the relative speed between gravitons and photons \cite{TheLIGOScientific:2017qsa},  eliminating part of the derivative couplings of the scalar field to the curvature \cite{Creminelli:2017sry,Sakstein:2017xjx,Ezquiaga:2017ekz,Baker:2017hug}.\footnote{As discussed in \cite{deRham:2018red}, this conclusion can be evaded if new physics appears at a scale parametrically smaller than the observed LIGO/Virgo frequencies.} 
{More recently, it has been pointed out that the theories beyond Horndeski belonging to the Gleyzes-Langlois-Piazza-Vernizzi (GLPV) class \cite{Gleyzes:2014dya,Gleyzes:2014qga} display a cubic graviton-scalar-scalar interaction that can mediate an observable decay of the graviton into  dark energy particles \cite{Creminelli:2018xsv}. The same vertex is responsible for an anomalous gravitational wave dispersion, when the speeds of scalar and gravitational wave differ. These two effects may become important at frequencies relevant for LIGO/Virgo observations,  leading to a very tight constraint on these theories.\footnote{Recently, positivity bounds for scalar theories coupled to gravity  \cite{Bellazzini:2019xts} have decreased significantly the regime of validity of the corresponding EFTs (e.g. for the cubic galileon
\cite{Nicolis:2008in}).  It would be interesting to study how these theoretical constraints affect general {theories with screening}.} 

In this article we study the subclass of scalar-tensor theories that leave gravitational waves unaffected and satisfy these constraints exactly.  This study is  the natural continuation of Ref.~\cite{Kobayashi:2014ida}, which examined the Vainshtein mechanism in GLPV theories, and of Refs.~\cite{Crisostomi:2017lbg,Langlois:2017dyl,Dima:2017pwp}, where this investigation was extended to Degenerate Higher-Order Scalar-Tensor (DHOST) theories \cite{Langlois:2015cwa,Crisostomi:2016czh,BenAchour:2016fzp}, in the case where the speed of gravitational waves equals the one of light.  
 These references showed that for the subclass of these theories extending Horndeski, the two gravitational 
potentials  differ inside matter and depend on the density profile, signalling the breaking of the Vainshtein screening. (See for instance also \cite{Bartolo:2017ibw,Ganz:2018vzg,Babichev:2018rfj,Kase:2018iwp,Crisostomi:2017pjs,Crisostomi:2018bsp,Frusciante:2018tvu} for other studies on the surviving theories and \cite{Kase:2018aps,Ezquiaga:2018btd} for reviews.)

As in these references, here we consider the solutions for the gravitational potentials in these theories near matter sources, i.e.~in the Vainshtein regime, and compare them to observational data.}  We will rely on the quasi-static approximation, which is valid on scales much smaller than the Hubble radius when we
restrict to non-relativistic sources.

In the next section  we review {DHOST theories}  and their effective description, and we expand the action in the metric potentials and scalar field perturbations. We then discuss the subset of theories that evade the gravitational wave constraints. We briefly study the linear theory in Sec.~\ref{sec:linear} and the Vainshtein regime around a spherically symmetric source in Sec.~\ref{vainshteinsec}.  Here, we also discuss our results in the more familiar Horndeski frame. {The previous discussions do not apply to theories that are related to General Relativity by a conformal transformation, which are the subject of Sec.~\ref{sec:xizero}.} Finally, constraints on the parameters are derived in Sec.~\ref{constraintssec}, and conclusions are left to Sec.~\ref{conclusion}. 

To simplify the main text, many of the coefficients and equations have been postponed to App.~\ref{sec:coeff}. In  \appref{otherconstraints} we also present some extra astrophysical constraints, and we show that they are weaker than the ones obtained from the  Solar System and the Hulse-Taylor pulsar.

\section{Action and perturbation equations}
\label{DHOST}

\subsection{DHOST theories}

We denote by the semicolon  a covariant derivative and we define $X \equiv - \phi_{;\mu} \phi^{;\mu}/2 $. The action for DHOST theories includes all possible quadratic combinations up to second derivatives of the field $\phi$   and reads \cite{Langlois:2015cwa}
\begin{align}
\begin{split}
\label{TypeIa}
S_{\rm DHOST} = \ &\int d^4 x \sqrt{-g} \Big[ P(\phi,X)+ Q(\phi,X) \Box \phi \\ 
&+ f(\phi,X) {}^{(4)}\!R + \sum_{I=1}^5 a_I (\phi,  X ) L_I (\phi, \phi_{;\nu},  \phi_{;\rho \sigma})  \Big] \;, 
\end{split}
\end{align}
where ${}^{(4)}\!R$ is the 4D Ricci scalar
 and the $L_I$ are defined by
\begin{align}
L_1&=\phi_{;\mu\nu}\phi^{;\mu\nu} \;, \quad  L_2=(\phi^{;\mu}_{;\mu})^2 \;, \quad  L_3=(\phi^{;\mu}_{;\mu})(\phi^{;\rho} \phi_{;\rho\sigma}\phi^{;\sigma})\;, \nonumber \\
  L_4&=\phi^{;\mu}\phi_{;\mu\nu}\phi^{;\nu\rho}\phi_{;\rho} \;, \quad  L_5=(\phi^{;\rho} \phi_{;\rho\sigma}\phi^{;\sigma})^2 \;.
\end{align}
The functions $P$ and $Q$ do not affect the degeneracy character of the theory. Instead, for DHOST theories,  the functions appearing in the second line of \eqn{TypeIa} must satisfy three degeneracy conditions \cite{Langlois:2015cwa} that fix three of these functions in terms of the others. Here we are going to focus on the subclass  that  satisfy $a_1+a_2 =0$ (and two other degeneracy conditions). Other subclasses have been shown to display a linear instability, either in the scalar or in the tensor sector {\cite{deRham:2016wji,Langlois:2017mxy}}.

\subsection{Effective description and constraints}

It is convenient to discuss observational constraints on these theories in terms of the EFT of dark energy parameters, which for DHOST theories have been introduced in \cite{Langlois:2017mxy} and extended to nonlinear order in the perturbations in \cite{Dima:2017pwp}.  

Specifically,  in the presence of a preferred slicing induced by a time-dependent  scalar field, we can choose the time as to coincide with the uniform field hypersurfaces. In this gauge, called the unitary gauge, and using the ADM metric decomposition with  line element  $ds^2 = - N^2 dt^2 + h_{ij} (dx^i + N^i dt) (dx^j + N^j dt)$, cosmological perturbations around an FRW solution  $ds^2=-dt^2 + a^2(t) d\vec x^2$ are governed by the action 
 \be
\begin{split}
\label{EFTaction}
&S_{\rm EFT} = \int  d^4 x \sqrt{h}  \frac{M^2}{2} \big[ -(1+\delta N)\delta {\cal K}_2 +c_T^2 {}^{(3)}\!R +4 H \alphaB \delta K \delta N \\
 &  +(1+\alphaH) {}^{(3)}\!R \delta N 
+4 \beta_1 \delta K V + \beta_2 V^2 + {\beta_3} a_i a^i
+ \alphaV \delta N \delta {\cal K}_2 \big]\;,
\end{split}
\ee
where we have written only the operators with the highest number of spatial derivatives, which are relevant {in the quasi-static limit.}
Here $H\equiv \dot a/a$ (a dot denotes the time derivative), $\delta N \equiv N-1$, $\d K_i^j \equiv K_i^j- H \delta_i^j$  is the perturbation of the extrinsic curvature of the time hypersurfaces, $\delta K$ its trace, and ${}^{(3)}\!R$ is the 3D Ricci scalar of these hypersurfaces. Moreover, $\delta {\cal K}_2 \equiv \delta K^2 - \delta K_{i}^j \delta K^{i}_{j}$, $V \equiv (\dot N - N^i \partial_i N)/N$ and $a_i \equiv \partial_i N/N$.

{The time-dependent functions in this action are related to the free functions in eq.~\eqref{TypeIa}. One finds that the effective Planck mass that normalizes the graviton kinetic energy is given by $M^2 = 2( f - 2 a_2 X)$. The other parameters read}   \cite{Dima:2017pwp}
\be
\label{alphabeta}
\begin{split}
\alphaB & = \alphaV -3 \beta_1 + \dot \phi ( f_\phi + 2 X f_{,\phi X} + X Q_{,X}) /(M^2H) \;, \\
c_T^2 &= {2 f}/{M^2}\,,  \\
  \alphaH & = 4 X (a_2 - f_{,X})/M^2  \,,
\\
\bun& =  2 {X} (f_{,X}- a_2+ a_3 X)/{M^2} \,, \\
\bdeux & =  -  {8X^2} \left(a_3+a_4  - 2 a_5X\right)/{M^2} \,, \\
\btrois &=-  {8X}( f_{,X} -  a_2-  a_4 X)/{M^2} \, ,\\
\alphaV & = {4 X ( f_{,X} - 2a_2 -2X a_{2,X} )}/{M^2} \;.
\end{split}
\ee
The function $\alphaB$ \cite{Bellini:2014fua} measures the kinetic mixing between metric and scalar fluctuations \cite{Creminelli:2008wc} and $c_T^2$  the fractional difference between the speed of gravitons and photons. The function $\alphaH$ measures the kinetic mixing between matter and the scalar fluctuations \cite{Gleyzes:2014dya,Gleyzes:2014qga,DAmico:2016ntq} and vanishes for Horndeski theories, while the function $\alphaV$ parameterizes the only operator that starts cubic in the perturbations \cite{Cusin:2017mzw}.

Finally,   
the functions $\beta_1$, $\beta_2$, and $\beta_3$ parameterize the presence of higher-order operators. {The  degeneracy conditions, which we assume hereafter, read }  \cite{Langlois:2017mxy}
\be
\label{degen}
\bdeux=-6\bun^2\,,\qquad   \btrois=-2\bun\left[2(1+\alphaH)+\bun c_T^2 \right] \;,
\ee
{so that only $\beta_1$ is independent.}
 Another  function that we define here for later convenience  is \cite{Bellini:2014fua}
\be
\alphaM \equiv \frac{d \ln M^2}{d \ln a}\; .
\ee

%
%
\subsection{Action in Newtonian gauge}

To study scalar linear and higher-order perturbations  we will exit the unitary gauge and work in the Newtonian gauge, where the metric is written as 
\be \label{newtongauge}
ds^2 = - ( 1 + 2 \Phi ) dt^2 + a^2(t) ( 1 - 2 \Psi ) d \vec{x}^2 \ .
\ee
The scalar field $\pi$ is introduced  by performing a space-time dependent shift in the time $t \rightarrow t + \pi ( t , \vec{x})$.  
Then, we expand the action \eqn{EFTaction} in terms of the metric and scalar field perturbations. We keep only {terms} with the highest number of derivatives per field, which are those relevant
in the quasi-static limit, and we find 
\be
\label{EFTactionexp}
S _{\rm EFT } = \int d^4 x  \left[ \frac{M^2 a }{2}  \left( \cL_2 + \frac{1}{a^2}\cL_3 + \frac{1}{a^4} \cL_4 \right) + \cL_{\rm m} \right] \, ,
\ee 
with 
\begin{align}
\begin{split}
\label{Lagrangians}
\cL_2   = \ & \left( c_1 \Phi + c_2 \Psi + c_3 \pi \right) \partial^2 \pi + c_4 \Psi \partial^2 \Phi + c_5 \Psi \partial^2 \Psi  \\ 
&  + c_6 \Phi \partial^2 \Phi + ( c_7 \dot \Psi + c_8 \dot \Phi + c_9 \ddot \pi ) \partial^2 \pi  \;, \\
\cL_3  = \ &  -\frac{b_1 }{2}  ( \partial \pi)^2 \partial^2 \pi  +   ( b_2 \Phi + b_3 \Psi ) Q_2 [\pi,\pi ]  \\
&  - \frac{1}{2 } ( \partial \pi)^2 ( b_4 \partial^2 \Psi + b_5 \partial^2 \Phi + b_6 \partial^2 \dot \pi )  \;, \\
\cL_4  = \ &  - \frac{d_1}{2}  ( \partial \pi )^2 Q_2[\pi,\pi]+ \frac{d_2}{4 } \partial_k ( \partial \pi)^2 \partial_k ( \partial \pi )^2     \;,
\end{split}
\end{align}
where {we have defined $Q_2 [ \varphi_a , \varphi_b]  \equiv \varepsilon^{ikm} \varepsilon^{jlm} \partial_i \partial_j \varphi_a \partial_k \partial_l \varphi_b $ for $\varphi_a \equiv \{\Phi, \Psi, \pi\}$, and} $c_1, \ldots, c_9$, $b_1, \ldots, b_6$ and $d_1, d_2$ are time-dependent coefficients, {reported} in \appref{coefficientsapp}. 

The last term in the bracket is the matter Lagrangian. If we define by $\rho_{\rm m}$ the matter energy density and by $ \bar \rho_{\rm m}$ its mean cosmological value, this is given by 
\be
\label{coupl}
\cL_{\rm m} =   - a^3 \Phi \delta \rho_{\rm m}  \ ,
\ee
where $\delta \rho_{\rm m} \equiv \rho_{\rm m} - \bar \rho_{\rm m}$.

The field equations can be derived straightforwardly by varying the action \eqn{EFTactionexp} with respect to $\Phi$, $\Psi$ and $\pi$.    We have
\be
\label{fieldeq}
{\cal E}_{\varphi_a} \equiv \frac{1}{2 a^2 M^2} \frac{\delta S_{\rm EFT} }{\delta \varphi_a}=0\;,
\ee
where the explicit expressions of ${\cal E}_{\varphi_a}$ {can be found in App.~\ref{coefficientsapp} (see eqs.~\eqref{phieq1}--\eqref{pieq1})}.

The above Lagrangians and equations  are valid for  general DHOST theories. We will now consider the subclass of theories leaving the  gravitational wave unaffected.

\subsection{Gravitational wave constraints}

{We now focus on the subset of theories that evade the gravitational wave constraints.
In particualr, we demand that gravity and light travel at the same speed, i.e.~(see e.g.~\cite{Creminelli:2017sry})},
\be \label{speed}
c_{\rm T}^2 = 1\ ,  \quad \text{and} \quad  \alpha_{\rm V} = - \alpha_{\rm H} \;, \qquad \text{(speed of gravity)} \ .
\ee
{Moreover, we require that gravitons do not decay into dark energy by setting} \cite{Creminelli:2018xsv}
\be \label{decay}
\alpha_{\rm H} +  2 \beta_1 =0 \;, \qquad \text{(no decay)} \, .
\ee
{Unless otherwise stated, in the following  we impose these two equations and replace $\alphaH$ and $\alphaV$ in terms of $\beta_1$.}

In terms of the  functions  $a_I$ in the action \eqn{TypeIa}, {these requirements, together with the degeneracy conditions, 
read $ a_1 = a_2 = a_3= a_5= 0 $ and $a_4 = 3 f_{,X}^2 /({2f})$,
for any $X$, i.e.,  \cite{Creminelli:2018xsv}} , 
\be
\begin{split}
\label{finaltheory}
S = \ &\int d^4 x \sqrt{-g} \Big[ P + Q \Box \phi + f  {}^{(4)}\!R +  \frac{3 f_{,X}^2}{2f} \phi^{;\mu}\phi_{;\mu\nu}\phi^{;\nu\rho}\phi_{;\rho}  \Big] \;.
\end{split}
\ee

{From eq.~\eqref{alphabeta}, in this subset of theories the remaing free parameters are thus given by}
\be
\begin{split}
\beta_1 &= X f_{,X} / f \;, \qquad \alphaM = f_{,\phi} \dot \phi / (f H)  \ , \\
\alphaB &=  - X f_{,X} / f + \dot \phi ( f_{,\phi} + 2 X f_{,\phi X} + X Q_{,X}) /(2f H) \;.
\end{split}
\ee

\section{Linear regime}
\label{sec:linear}

{We briefly discuss how matter inhomogeneities  source $\pi$ and the gravitational potentials at linear order. For convenience, we  first  define the combination
\be
\begin{split}
 \alpha  c_s^2  \equiv \ & {2}  ( 1 + \alphaB - \dot \beta_1 / H )^{2}   \bigg[  \frac{1}{a M^2} \frac{d}{dt }   \left( \frac{a M^2 (1-\beta_1)}{H ( 1 + \alphaB) - \dot \beta_1}  \right)  -1  \bigg]  \\
 &   \qquad - \frac{\bar \rho_{\rm m} ( 1 - \beta_1)^2}{ H^2 M^2} \;,
\end{split}
\ee
where $c_s^2$ is the
effective sound speed of dark energy fluctuations \cite{Langlois:2017mxy},
which must be positive to avoid gradient instabilities, while the time-dependent function $\alpha$ is the coefficient in front of the time kinetic term of scalar fluctuations (see \cite{Langlois:2017mxy}), which must be positive to avoid ghosts. In practice, we do not need their explicit expressions because} in the quasi-static limit these two parameters always appear in the combination $ \alpha c_s^2>0$. 

{By} solving for $\Phi$ and $\Psi$ the linear equations obtained by varying the action with respect to the two metric potentials, and replacing these solutions in  the scalar field equation, we obtain
\be
\label{linearpi}
\partial^2 \pi  = - \frac{a^2}{2 \mpl \nu_2} \left( \nufive \delta \rho_{\rm m} + \nusix  \delta  \dot \rho_{\rm m}  \right) \;,
\ee
where $\Mp$ is the value of the Planck mass measured  today and the parameters $\nutwo$, $\nufive$ and $\nusix$ above are defined as
\be
\begin{split} \label{nudef1}
\nutwo  &\equiv   \frac{ M^2 H^2   \alpha \,  c_s^2}{2 \Mp(1 - \beta_1)} \;, \\
\nufive & \equiv  \frac{-  H[   \alpha_{\rm B} -  \alpha_{\rm M} ( 1- \beta_1 ) + \beta_1 (4 - 3 \beta_1)] }{  1 - \beta_1} \, , \\
 \nusix  & \equiv -     \beta_1 \;.
\end{split}
\ee
This choice of definitions  will become clearer when we consider the full nonlinear equation for $\pi$, in \secref{vainshteinsec}.  

For completeness, we also provide the solutions for the metric potentials, which can be obtained by replacing \eqn{linearpi} back into the  equations for $\Phi$ and $\Psi$. One  finds  \cite{Crisostomi:2017pjs, Hirano:2019nkz}
\begin{align}
\begin{split} \label{linearsolutions}
\partial^2 \Phi&  = \mu_\Phi \delta \rho_{\rm m} + \nu_ \Phi {\delta \dot \rho}_{\rm m} + \sigma_\Phi {\delta \ddot \rho}_{\rm m} \ , \\
\partial^2 \Psi &= \mu_\Psi \delta \rho_{\rm m} + \nu_ \Psi {\delta \dot \rho}_{\rm m} + \sigma_\Psi {\delta \ddot \rho}_{\rm m} \ ,
\end{split}
\end{align}
where the coefficients on the right-hand side are given in \appref{coefficientsapp2}.  Observational constraints on the above coefficients of the linear equations are discussed in \cite{Hirano:2019nkz}.  
We are now ready to discuss the Vainshtein regime.

%
%
%
\section{Vainshtein regime} \label{vainshteinsec}

In this section we want to study the Vainshtein mechanism around a spherically symmetric body, such as for instance a non-relativistic star {({see \cite{Kobayashi:2018xvr} for a study of relativistic stars in DHOST theories}).} {This is at play close to the body, where non-linearities of the scalar field become important  suppressing the scalar force. Far from the body, the linear solutions affected by the fifth force (see \eqn{linearsolutions}) are recovered.  This can allow the theory to be compatible with stringent Solar System observations while at the same time modifying gravity on large scales. } 

To derive the equations relevant for spherically symmetric solutions, we assume that all of the fields depend only on time and the radial variable, $\varphi_a ( t , \vec x) = \varphi_a ( t , r )$, where $r \equiv | \vec x|$.  Then, we integrate the field equations \eqn{fieldeq} over the radial variable and use Stoke's theorem. Following \cite{Kobayashi:2014ida}, we use the following notation,
\be \label{xyzdefs}
x \equiv \frac{1}{\Lambda^3} \frac{\pi'}{a^2 r} \ , \qquad y \equiv \frac{1}{\Lambda^3} \frac{\Phi'}{a^2 r} \ , \qquad  z \equiv \frac{1}{\Lambda^3} \frac{\Psi'}{a^2 r} \ ,  
\ee
where a prime denotes a derivative with respect to $r$ and  $\Lambda$ is a mass scale of order $\Lambda^3 \sim H_0^2 \Mp$, where $H_0$ is the Hubble rate today.  Moreover, we define
\be
\label{mdef}
 \mathcal{A} \equiv \frac{1}{8 \pi \Mp \Lambda^3} \frac{m}{a^3 r^3} , \qquad  m  \equiv 4 \pi \int_0^r d \tilde r \, \tilde r^2 \, a(t)^3  \delta \rho_{\rm m} ( t , \tilde r) \;,
\ee 
where $m$ is the {\em physical} mass of an overdensity contained in a spherical ball of physical radius $a(t)r $. 
The quantity ${\cal A}(t,r)$ represents the {\em comoving} mass density contrast in  this  ball.

The explicit expressions of the field equations in spherical symmetry are reported in \appref{coefficientsapp3}.
Following  \cite{Crisostomi:2017lbg,Langlois:2017dyl,Dima:2017pwp}, we can solve the first two equations for $y$ and $z$ and plug the solutions into the third equation. 
After imposing the degeneracy conditions \eqn{degen},  one obtains a  polynomial equation for $x$ only. This equation was previously studied after imposing  that gravitons and photons propagate with the same speed, \eqn{speed}, in \cite{Crisostomi:2017lbg,Langlois:2017dyl} and in the general case in \cite{Dima:2017pwp} and a cubic equation was obtained. Here we further restrict to theories where the gravitational waves do not decay, i.e.~\eqn{decay}. In this case one finds instead a {\em quadratic} equation, 
\be  \label{scalarspher}
\nuone x^2 + \left[ \nutwo  +  \nufour r^{-2} (r^3 \mathcal{A})' \right]x + \nufive  \mathcal{A} +\nusix  \mathcal{ \dot A} = 0 \; ,
\ee
where the time-dependent functions $\nutwo$, $\nufive$  and $\nusix$  {already appeared in the linear equations and are} defined in \eqn{nudef1}. The other functions are given by
\be
\begin{split} \label{nudef}
\nuone  & =   \xi \mpl^{-1} M^2 H \Lambda^3 \ , \quad \quad \nufour  = \Lambda^3 \beta_1 \;,
\end{split}
\ee
with 
\be
\label{xidef}
 \xi \equiv  2 \alphaB  - \alphaM (1 - \beta_1) + 2 \beta_1    - 2 \dot \beta_1 / H  \;.
 \ee

{If $\nu_1 \neq 0$,  the term proportional to $x^2$ becomes important close to the matter source, while it is negligible far away from it. The transition happens at the so-called Vainshtein radius, roughly corresponding to $\mathcal{A} \sim 1$, i.e.,
\be
 r_V \sim \left( \frac{m}{8 \pi M_{\rm Pl}} \right)^{1/3} \frac{1}{a \, \Lambda}  \ . 
\ee 
For example, for a star like the sun, the Vainshtein radius is about a tenth of the size of the Milky Way. 
}

{For $\nu_1 = 0$, the term quadratic in $x$ vanishes and the usual Vainshtein screening cannot be supported. As shown in  \secref{sec:Frame}, this case corresponds to scalar-tensor theories that are conformally related to General Relativity, i.e., that can be described by the Einstein-Hilbert term plus conformally coupled matter. Here we will assume that $\nuone$ (and thus $\xi$) does not vanish. We discuss the case $\xi = \nuone =0$ in Sec.~\ref{sec:xizero}.
}

As discussed in Sec.~\ref{sec:linear}, $\alpha c_s^2>0$. Thus, for $\beta_1<1$, $\nutwo$ defined in \eqn{nudef1} is also positive and the solution to \eqn{scalarspher} that matches the linear regime, i.e., that has the correct behavior for $\mathcal{A} \ll 1$,  is 
\begin{align}
\begin{split}
\label{fullsolution}
x & = -  \frac{  \nutwo  +  \nufour \kappa {\cal A}  - \sqrt{ \left( \nutwo  +  \nufour \kappa {\cal A} \right)^2 - 4 \nuone (\nufive  \mathcal{A}  + \nusix \dot{ \mathcal{A} })  } }{2 \nuone} \ ,
\end{split}
\end{align}
where for convenience we have defined the positive function, 
\be
\kappa (t,r) \equiv  \frac{\partial \ln m (t,r)}{\partial \ln r}  \geq 0\;.
\ee

The terms proportional to $\nufour$ and $\nusix$ in \eqn{scalarspher} contain respectively the radial and time derivative  of the  comoving mass of the body.  Thus, when the mass of the central overdensity is constant in time and space, such as at a radial distance larger than the body size, we have $(r^3 \mathcal{A})' =0$ and $\dot{\mathcal{A} }= - 3 H \mathcal{A}$. Therefore, we consider two different cases, depending on whether we are outside or inside the object.

%
\subsection{Inside of matter} \label{insidesec}

Inside of matter, $(r^3 \mathcal{A})'$ is generally different from zero,
in which case, when $\mathcal{A} \gg 1$, the term proportional to $\nufour$ can dominate the square root in \eqn{fullsolution}.  This regime is characterized by 
\be \label{insideconditions}
\mathcal{A} |\beta_1|  \kappa \gg  \frac{\beta_1 \nutwo}{\nufour}  \sim \frac{\alpha  c_s^2}{2 (1-\beta_1)}\ ,
\ee
and
\be \label{insideconditions1}
\mathcal{A} \left( \beta_1 \kappa\right)^2 \gg \frac{ 4 \beta_1^2 \nuone (\nufive - 3 H \nusix) }{\nufour^2 }  \sim  \frac{4 \xi (\nufive - 3 H \nusix)}{H}  \, ,
\ee
where on the right-hand side we have used the definitions in \eqn{nudef1} and \eqn{nudef} and that $M^2 H^2 \sim \Mp \Lambda^3$.
The right-hand sides of these equalities are  of order unity while, on the left-hand side, ${\cal A} \gg 1$.  For instance, inside a star like the sun,\footnote{Here and in the rest of this paper, we use the symbol $\odot$ to denote the appropriate quantity for our sun.}  
\be
{\cal A} \sim 10^{30} \frac{m}{M_\odot} \left( \frac{R_\odot }{ r }  \right)^3 \;,
\ee
where we have used $\Lambda \sim (10^3 \, \text{km})^{-1}$.
These conditions are thus realized inside matter unless $\beta_1 \kappa $ becomes  $\mathcal{O}(10^{-15})$.  {Using a mass profile from the Lane-Emden equation (see \eqn{leeq} in the appendix), we have verified that this happens only on a very thin region near the surface of the star.  } 

Now, expanding the solution in \eqn{fullsolution} for $\kappa \neq 0$ and $\mathcal{A} \gg 1$, we can distinguish between two cases. 
When $\beta_1>0$, the solution for $x$ is  given by
\begin{align} \label{xsolinside}
x_{\rm in}   \approx - \frac{\nufive - 3 H \nusix    }{\nufour \kappa   }  \ , \qquad (\beta_1> 0 \;, \  \mathcal{A} \gg 1)\;.
\end{align}
This solution can then be plugged back into the field equations for $y$ and $z$ (see 
eqs.~\eqref{potspher1} and \eqref{potspher2}) to solve for these two variables.  Since $x_{\rm in} \sim {\cal O} ({\cal A}^0)$, the equations for $y$ and $z$ are dominated by the usual matter term linear in ${\cal A}$ and
terms both linear and quadratic in $x$  can be neglected. 

The solutions for the potentials can be straightforwardly computed and read 
\begin{align} \label{leadinginside} 
\Phi_{\rm in}'  =  \frac{G_* (1+\varepsilon_\Phi^{\rm in}) m }{r^2a } \;, \qquad \Psi_{\rm in}'  =  \frac{G_* (1+\varepsilon_\Psi^{\rm in}) m }{r^2a }   \ ,
\end{align}
where $G_* \equiv 1/(8 \pi M^2)$ is the  gravitational constant that canonically normalizes the graviton, and  
\begin{align}
\label{insidevarespilon}
\epsilon_\Phi^{\rm in}  \equiv  \frac{\beta_1( 2 -\beta_1)}{(1-\beta_1)^2}\ ,  \qquad \epsilon_\Psi^{\rm in} & \equiv    \frac{-\beta_1^2}{(1-\beta_1)^2} \ . 
\end{align}
Thus,  in this theory $\Phi \neq \Psi$ so that Vainshtein screening is broken inside the body. 
Notice that the breaking is different from the one found in \cite{Kobayashi:2014ida,Crisostomi:2017lbg,Langlois:2017dyl,Dima:2017pwp}, which depends on the radial derivatives of the object mass.

The case $\beta_1<0$ is instead ruled out.  Indeed, in this case  the solution  \eqn{fullsolution}  reads
\be
x_{\rm in}  \approx \frac{|\nufour| \kappa}{\nuone} {\cal A} \ , \qquad (\beta_1< 0 \;, \ \mathcal{A} \gg 1) \;.
\ee
Therefore, $\pi$ is not suppressed inside matter and nonlinear terms proportional to $x^2 \sim {\cal O}({\cal A}^2)$ dominate the field equations for $y$ and $z$. {The solutions,} 
\be
\Phi_{\rm in}'\approx -\Psi_{\rm in}' \approx  \frac{\beta_1^3(r \kappa'+ \kappa^2 - 2 \kappa)}{(1-\beta_1) r} \left( \frac{ G_* m }{ H r^2 a^2}   \right)^2 \;,
\ee
{are} incompatible with the existence of stars or other bounded objects.

%
\subsection{Outside of matter} \label{outsidesec}

Using \eqn{mdef} outside the object, where the physical mass is constant, we have $ \dot{ \mathcal{A}} = - 3 H \mathcal{A}$ and $ (r^3 \mathcal{A})' = 0$. Replacing this in \eqn{fullsolution},  near the source  we obtain 
\be \label{mpzero}
x_{\rm out} \approx  \frac{\sqrt{ -  \nuone  (\nufive - 3 H \nusix)  \mathcal{A} } }{\nuone} \ , \qquad (\mathcal{A} \gg 1 ) \;.
\ee
In contrast with the solution inside matter, \eqn{xsolinside}, where  $x_{\rm in} \sim {\cal O} ({\cal A}^0)$,   here  $x_{\rm out}^2 \sim {\cal O}({\cal A})$ so that terms quadratic  in $x$ contribute to the $y$ and $z$ solutions while linear terms are negligible. 

The solutions for the potentials become now   
\begin{align} \label{yzsols}
\Phi_{\rm out}'  =  \frac{G_* (1+\varepsilon_\Phi^{\rm out}) m }{r^2 a} \;, \qquad \Psi_{\rm out}'  =  \frac{G_* (1+\varepsilon_\Psi^{\rm out}) m }{r^2 a }   \ ,
\end{align}
with  
\begin{align}
\begin{split} \label{epsout}
\varepsilon_{\Phi}^{\rm out}   \equiv  \ & \varepsilon_{\Phi}^{\rm in}  - \frac{\beta_1 \upsilon}{2 \xi (1 - \beta_1)^2} \ , \\
\varepsilon_{\Psi}^{\rm out}   \equiv \ &  \varepsilon_\Psi^{\rm in} + \frac{\beta_1 \upsilon}{2 \xi (1 - \beta_1)^2}   \ ,
\end{split}
\end{align}
where for convenience we have defined a new quantity,
\be
\label{upsilondef}
\upsilon \equiv \alphaB - \alphaM(1-\beta_1) + \beta_1 \ . 
\ee
From these expressions, it is clear that the Vainshtein mechanism is broken also outside of the matter source because the two gravitational potentials are different. 
Only for
\be
\label{tuning}
2 \xi = \upsilon \;
\ee
are the two potentials the same and the Vainshtein screening recovered outside of matter.

\begin{figure}
\centering
\includegraphics[width=8.6cm]{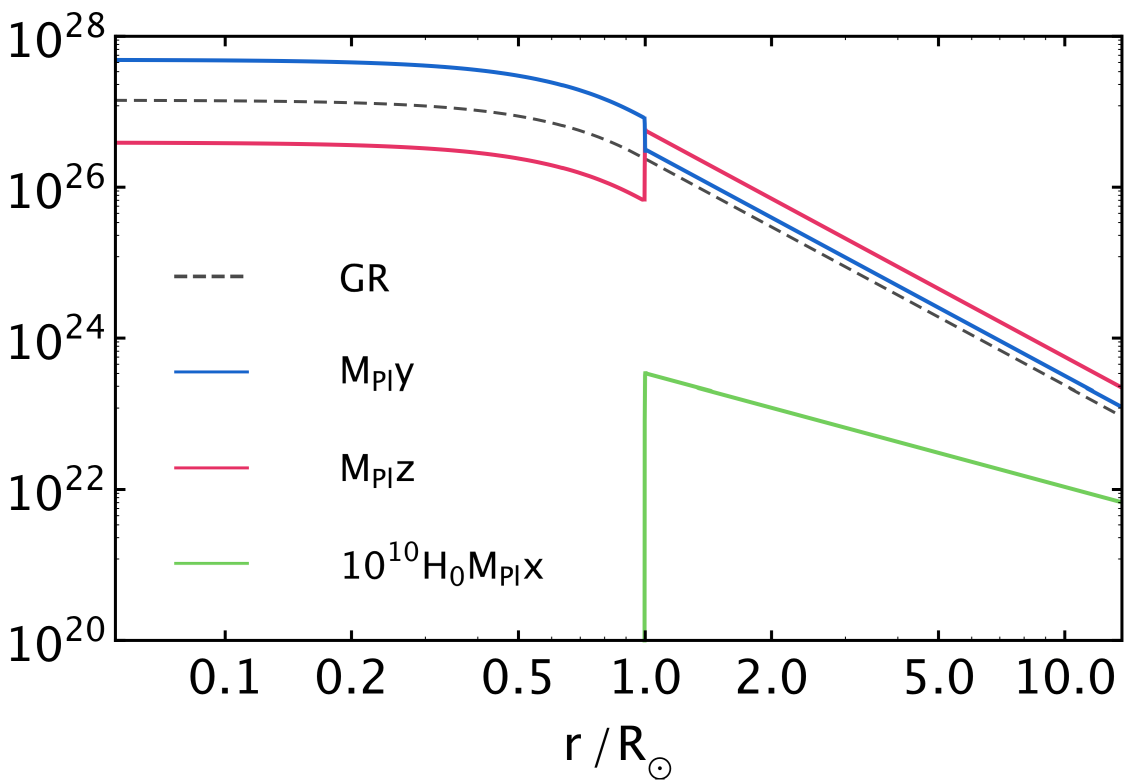}
\caption{ {An} example of the solutions for the potentials inside and outside of a matter source which has the same mass and radius as the sun.  {For the mass profile of the star, we used the solution to the Lane-Emden equation, \eqn{leeq}.}  {We chose large parameters, i.e.,} $\alphaB = -0.4766$, $\alphaM = -0.08$, and $\beta_1 = 0.46$, to emphasize the deviations from General Relativity.  Notice that, to fit it on the plot, we had to scale the solution for $x$ by a factor of $10^{10}$.  Since $\mathcal{A} \sim 10^{30}$, the transition between outside and inside solutions is abrupt, and the solution for $x$ is highly suppressed, with $H_0 \mpl x \sim \mathcal{O}(1)$ inside of the star.     }
\label{potplot}
\end{figure}

{An example of the  solutions \eqn{leadinginside} (for $r< R_\odot$) and \eqn{yzsols} (for $r> R_\odot$) is presented in \figref{potplot}. Notice the suppressed value of $x$. For a large overdensity ${\cal A}$, such as in a star, the transition between the interior of the star, where eqs.~\eqref{insideconditions} and \eqref{insideconditions1} are satisfied, and the exterior, where $\kappa=0$, takes place abruptly, so that $x$, $y$ and $z$ visually display a discontinuity for $\beta_1 \upsilon \neq 0$. Of course, the solutions are actually continuous, as follows for instance from \eqn{fullsolution}.}

%
%
%
%

%
%
%
%
%
\subsection{The Horndeski frame}
\label{sec:Frame}

The above results can be understood by noticing   
 that DHOST theories that we consider in this paper can be mapped to Horndeski theories via an invertible $X$-dependent conformal and disformal transformation  \cite{Crisostomi:2016czh,Achour:2016rkg}.  
 
 Since we  want to preserve the speed of propagation of gravitons,  we  focus on  conformal transformations only, i.e. 
\be
\label{change_frame}
g_{\mu \nu} \to \tilde g_{\mu \nu}= C(\phi, X) g_{\mu \nu} \; , 
\ee
and use the tilde to denote quantities in the Horndeski frame.
For convenience, we also introduce the dimensionless time-dependent parameter 
\be
\alpha_{\rm Y} = - \frac{d \ln C}{d \ln X} \;,
\ee
with the right-hand side of this equation  evaluated on the background solution.  
 
The relations to compute the transformation of the EFT parameters of the  action \eqn{EFTaction} under the  change of frame \eqn{change_frame} are given in \cite{Langlois:2017mxy} (see for instance eq.~(2.22) of that reference). Without loss of generality, we will assume that $C=1$ on the background solution, so that $\tilde M^2 = M^2$ and $\tilde a =a$. Moreover, for the transformation above, one finds
\be
\begin{split}
\tilde \alpha_{\rm H}  = \frac{\alphaH - 2 \alpha_{\rm Y}}{1+\alpha_{\rm Y}}  \;, \qquad
\tilde \beta_1 = \frac{\beta_1 + \alpha_{\rm Y}}{1+\alpha_{\rm Y}} \;,
\end{split}
\ee
while $\tilde c_T = c_T$. 

Using these relations, one can show  that
a DHOST theory {with gravitational metric $g_{\mu \nu}$  satisfying the conditions \eqn{speed} and \eqn{decay} (see its covariant form in eq.~\eqref{finaltheory}), is equivalent to a Horndeski theory  with gravitational metric
$\tilde g_{\mu \nu}$,\footnote{Since matter is coupled to $g_{\mu \nu}$,  the coupling to matter in the new frame $\tilde g_{\mu \nu}$ is   non-minimal; see below.} with
$\tilde c_T=1$ and $\tilde  \alpha_{\rm V}= \tilde  \alpha_{\rm H}= \tilde  \beta_1 = 0 $ and action
\be
\begin{split}
\tilde S[\phi, \tilde g_{\mu \nu}] = \ &\int d^4 x \sqrt{-\tilde  g} \Big[ {}^{(4)}\!\tilde R + \tilde P(\phi, \tilde X) + \tilde  Q (\phi, \tilde X)\Box \phi   \Big] \;,
\end{split}
\ee
provided that $\alpha_{\rm Y} = - \beta_1$.} Additionally, we obtain  \cite{Langlois:2017mxy} (see eq.~(C.14) of that reference)
\be \label{alphabtrans}
\tilde \alpha_{\rm B}  = \frac{   \alphaB + \beta_1-  \dot \beta_1/H}{1- \beta_1  }  \;, \qquad \tilde \alpha_{\rm M} = \alphaM \, .
\ee

Using the above transformations we can check that $\xi$ given in \eqn{xidef} can be written in terms of Horndeski-frame quantities as $\xi = (1-\beta_1) (2 \tilde \alpha_{\rm B} -\tilde \alpha_{\rm M} ) $. As mentioned below \eqn{xidef}, this vanishes for $\tilde \alpha_{\rm M} = 2 \tilde \alpha_{\rm B}$, i.e.~for Brans-Dicke \cite{Brans:1961sx}, $f(R)$  \cite{Carroll:2004de} or other theories conformally related to General Relativity \cite{Gleyzes:2015pma}.

In the Horndeski frame, the  EFT action  \eqn{EFTactionexp} is given by
\be
\label{EFTactionexp2}
\tilde S _{\rm EFT } = \int d^4 x  \left[ \frac{M^2 a }{2}  \left( \tilde \cL_2 + \frac{1}{a^2}\tilde \cL_3 + \frac{1}{a^4} \tilde  \cL_4 \right) + \tilde \cL_{\rm m} \right] \, .
\ee 
Here the Lagrangians $\tilde {\cal L}_2$, $\tilde {\cal L}_3$ and $\tilde {\cal L}_4$  have  analogous expressions as those in \eqn{Lagrangians}, but now fields and parameters are  tilded and $\tilde \beta_1=\tilde \beta_2=\tilde \beta_3=0$.

Using \eqn{change_frame} and focusing on the leading order in spatial derivatives to retain only terms relevant in the quasi-static limit, the potentials in the Horndeski frame are related to those in the DHOST frame by 
\begin{align}
\begin{split} \label{phipsitransf}
\tilde \Phi & = \Phi - \beta_1  \left[ \Phi - \dot \pi  + \frac{1}{ 2a^{2}} ( \partial \pi)^2 \right] \ , \\
\tilde \Psi & = \Psi + \beta_1  \left[   \Phi - \dot \pi + \frac{1}{ 2a^{2}} ( \partial \pi)^2 \right] \ , 
\end{split}
\end{align}
while it is straigforward to verify that the scalar field  fluctuation does not change,
\be
\tilde \pi  = \pi  \;.
\ee

While the metric transforms when changing frame, matter remains always minimally coupled to the gravitational metric in the DHOST frame, i.e.~$g_{\mu \nu}$. 
Therefore, $\tilde {\cal L}_{\rm m}  = {\cal L}_{\rm m}$. Using \eqn{coupl} and the relations above,  the coupling in the Horndeski frame is thus
\be \label{hornmatter}
\tilde{ \mathcal{L}}_{\rm m}  = -  \frac{a^3}{1 - \beta_1} \left[ \tilde \Phi - \beta_1 \left( \dot {\tilde \pi} - \frac{1}{2 a^2} ( \partial \tilde \pi)^2 \right) \right] \delta  \rho_{\rm m} \ .
\ee

We can now vary the action \eqn{EFTactionexp2} with respect to $\tilde \Phi$, $\tilde \Psi$ and $\tilde \pi$. Actually, 
using the above functional relationships between the Horndeski action $\tilde S_{\rm EFT}$ and the DHOST action $S_{\rm EFT}$, we can easily derive the relationship between the Horndeski frame equations of motion, $\tilde{\mathcal{E}}_{\rm \tilde \varphi_a}  \equiv (2 a^2 M^2)^{-1} \delta \tilde S_{\rm EFT} / \delta \tilde \varphi_a=0$ and the DHOST frame equations of motion \eqn{fieldeq} using the chain rule,
\be
\frac{\delta S_{\rm EFT}}{\delta \varphi_a} = \sum_b \frac{ \delta \tilde \varphi_b}{\delta \varphi_a} \frac{\delta \tilde S_{\rm EFT}}{ \delta \tilde \varphi_b}  \ .
\ee
This gives
\begin{align}
\begin{split}
\tilde{ \mathcal{E}}_{\tilde \Phi} & =  \mathcal{E}_\Phi + \beta_1 (1 - \beta_1)^{-1} ( \mathcal{E}_\Phi -  \mathcal{E}_\Psi ) \ , \\
\tilde{\mathcal{E}}_{\tilde \Psi} & = \mathcal{E}_\Psi \ , \\
\tilde{\mathcal{E}}_{\tilde \pi} & = \mathcal{E}_\pi + \frac{1}{a M^2} \partial_t \left[ \frac{a M^2 \beta_1 \left( \mathcal{E}_\Phi - \mathcal{E}_\Psi \right)}{1 - \beta_1} \right]  \\
&\qquad \qquad \qquad \qquad    - \frac{\beta_1}{a^2} \partial_i \left[ \frac{\left( \mathcal{E}_\Phi - \mathcal{E}_\Psi \right) \partial_i \pi  }{1 - \beta_1}\right] \, .
\end{split}
\end{align}

Using the  field equations $\tilde {\cal E}_{\tilde \varphi_a}=0$, we can 
solve the system assuming spherical symmetry around a body,  similarly to what was done earlier in \secref{vainshteinsec}. 
The equations for $\tilde y$ and $\tilde z$ read
\be
-H \tilde \alpha_{\rm B} \tilde x +  \tilde z = \frac{ \mathcal{A}}{M ( 1 - \beta_1)} \;, \qquad H \tilde \alpha_{\rm M} \tilde x +  \tilde y = \tilde z  \;,
\ee
(i.e. all nonlinear terms vanish), and the closed equation for $\tilde x$ remains the same as that of $x$, \eqn{scalarspher}, as expected.  Since terms linear in $\tilde x$ can be neglected both inside and outside of matter,  
the solutions to the above equations  become, in terms of the potentials in the Horndeski frame,
\be \label{simplesol}
\tilde \Phi'  = \tilde \Psi' = \frac{G_*}{ 1 - \beta_1}  \frac{m }{r^2 a }\ ,
\ee
both inside and outside of the source, as long as ${\cal A} \gg 1$.

With \eqn{simplesol}, we can {now verify what we found in the previous subsections.}  Inside matter, $\pi$ can be neglected in \eqn{phipsitransf} and \eqn{leadinginside} is recovered.  Outside matter, $\dot \pi$ can be neglected but $(\partial  \pi)^2$ cannot. Replacing  in \eqn{phipsitransf} the solution for $x$, \eqn{mpzero}, we can recover the solution outside the body, \eqn{yzsols}.

{Moreover, the conformal transformation \eqn{phipsitransf} leaves the sum of the potentials invariant. 
One can verify directly that our expressions inside \eqn{leadinginside}  and outside \eqn{yzsols} of the source satisfy 
$y + z = \tilde y + \tilde z = 2 G_* m / (  r^3 \Lambda^3  (1 - \beta_1)) $, where we have used \eqn{simplesol} for the last equality.}
Additionally, the fact that the solutions in the Horndeski frame \eqn{simplesol} are {valid both} inside or outside of the source means that $\epsilon_{\Phi}^{\rm out} + \epsilon_{\Psi}^{\rm out} = \epsilon_{\Phi}^{\rm in} + \epsilon_{\Psi}^{\rm in}$, which can be verified directly in \eqn{epsout}.

\section{Theories conformally related to General Relativity}
\label{sec:xizero}

In the previous section we focused on theories with $\xi \neq 0$. Here we consider the case   $\xi = 0$, which corresponds to theories  related to General Relativity by the conformal transformation \eqref{change_frame}, as shown in Sec.~\ref{sec:Frame}.
Their general action is given by \eqn{finaltheory} with $Q = 0$. Examples are Jordan-Brans-Dicke \cite{Brans:1961sx} and $f(R)$ theories \cite{Carroll:2004de},  {but here we are interested in extentions of these theories where the conformal factor relating them to General Relativity depends also on $X$, in which case $\beta_1 \neq 0$.}

When $\xi = 0 = \nu_1$, the equation for $x$, \eqn{scalarspher}, becomes linear.  Inside of matter, the term proportional to $\nu_3$ dominates the solution so that one obtains the same conclusions as in \secref{insidesec} (including the constraint $\beta_1 \geq 0$).  

Thus,  we focus on the solutions outside of matter and near the source.  In that case, the solution for $x$ is simply the linear solution,
\be
x_{\rm out} = - \frac{( \nu_4 - 3 H \nu_5 ) \mathcal{A}}{\nu_2} = \frac{2 \Mp \upsilon}{\alpha c_s^2 H M^2} {\cal A} \ .
\ee
Solving the equations for $y$ and $z$ with this solution
we have, in terms of $\Phi'$ and $\Psi'$, 
\be
\label{novain}
\Phi_{\rm out}' \approx - \Psi_{\rm out} ' \approx  \frac{8\beta_1 \upsilon^2}{c_s^4 \alpha^2 (1-\beta_1) r} \left( \frac{ G_* m }{ H r^2 a^2}   \right)^2 \;.
\ee
This is clearly incompatible with Solar System tests unless $\beta_1 \upsilon^2$ is practically zero. 

A possible way out {with $\beta_1 \neq 0$ is to consider the subclass of theories with $\upsilon=0$.} Taking into account that $\xi=0$, this fixes all parameters as a function of $\beta_1$, i.e.,
\be
\label{coeffalphas}
\alphaB = \frac{2 \dot \beta_1}{H} - \beta_1\;, \qquad  \alphaM = \frac{2 \dot \beta_1}{(1-\beta_1) H} \;.
\ee
For this particular case the source term in \eqn{scalarspher} is absent and the fifth force vanishes, $x = 0$. Thus, the gravitational potentials outside the source are linear in the object mass and coincide with the solutions \eqn{leadinginside}, which were obtained ignoring the scalar field contribution.

%
%
%
\section{Constraints} \label{constraintssec}

{Let us discuss observational constraints on the parameters of the theories studied above, in particular on $\beta_1$.
A first class of  constraints comes from stellar physics. We have studied them in App.~\ref{otherconstraints} and have shown that they lead to complementary (but weaker) constraints to those derived in the following.}

As shown in \cite{Jimenez:2015bwa}, for $c_T=1$ the decrease of the orbital period of binary stars is proportional to the ratio between the gravitational constant normalizing the gravitons, $G_*$, and the one entering the Kepler law, which here is taken to be the one outside the object, $G_* (1 + \epsilon^{\rm out}_\Phi)$.
Thus, using the results from \cite{Weisberg:2010zz}, the Hulse-Taylor pulsar (PSR B1913+16) \cite{Hulse:1974eb} allows us to constrain $\epsilon_\Phi^{\rm out}$, 
\be \label{htconst}
- 2.5 \times 10^{-3} \le  \epsilon_\Phi^{\rm out} \le    7.5 \times 10^{-3} \, \text{  at  } \, \,  2\sigma.  
\ee

Next, we move to the Cassini constraints.  Measurements of the frequency shift of radio waves, sent to and from the Cassini spacecraft as they passed near the sun \cite{Bertotti:2003rm}, constrain the post-Newtonian parameter $ \gamma_{\rm PN}   \equiv \Psi  / \Phi$ to be $-0.2 \times 10^{-5}<\gamma_{\rm PN} - 1  < 5.5  \times10^{-5}$.  
Since \eqn{htconst} says that $\epsilon_\Phi^{\rm out}$ is small, we can approximate $\gamma_{\rm PN} - 1 \approx \epsilon_\Psi^{\rm out} - \epsilon_\Phi^{\rm out} $ and  use this measurement to constrain the relative difference between the gravitational potentials outside of matter, 
\be \label{cassiniconst}
-0.2 \times 10^{-5}<  \epsilon_\Psi^{\rm out} - \epsilon_\Phi^{\rm out}  < 5.5  \times10^{-5} \ . 
\ee

Using \eqn{yzsols} we can rewrite the above quantity as
\be \label{gammaexp}
\epsilon_\Psi^{\rm out} - \epsilon_\Phi^{\rm out} = \frac{\beta_1 ( \upsilon - 2 \xi)}{\xi (1-\beta_1)^2} \ ,
\ee
where we remind the reader that $\xi$ and $\upsilon $ are respectively defined in eqs.~\eqref{xidef} and \eqref{upsilondef}.
For generic values of $\alphaB$, $\alphaM$ and $\dot \beta_1$, one expects that $2 - \upsilon/ \xi  \sim \mathcal{O}(1)$, so that the above turns into a tight constraint on $\beta_1$,
\be
0 \le \beta_1  \lesssim 10^{-5}\; .
\ee
The bound on the left-hand side comes from  the result of \secref{insidesec} that negative values of $\beta_1$ are not allowed.

A more conservative constraint on $\beta_1$, independent of $\alphaB$, $\alphaM$, and $\dot \beta_1$,  comes from isolating $\beta_1$ by using \eqn{epsout}. One obtains
\be
\frac{\beta_1}{1 - \beta_1} = \epsilon^{\rm out}_\Phi - \half ( \epsilon^{\rm out}_\Psi - \epsilon^{\rm out}_\Phi ) \ . 
\ee
Combined with the constraint $\beta_1>0$, this gives 
\be
0 \leq \beta_1 <  7.5 \times 10^{-3} \, \text{  at  } \, \,  2\sigma  \ ,
\ee
which also holds in the presence of the tuning $2\xi- \upsilon \approx 0$. 

{The previous discussion assumed $\xi\neq 0$. Theories that are conformally related to General Relativity have $\xi=0$. As discussed in Sec.~\ref{sec:xizero}, these theories are ruled out unless they satisfy the condition \eqref{coeffalphas}.
In this case, the post-Newtonian parameter $\gamma_{\rm PN}$ reads
\be
\gamma_{\rm PN} -1 \approx - \frac{2 \beta_1}{(1 - \beta_1)^2} \ .
\ee
Combining $\beta_1 \geq 0$ with the Cassini constraint, \eqn{cassiniconst},  gives 
\be
0 \leq \beta_1 \lesssim 10^{-6} \ . 
\ee
}

%


\section{Conclusion} 
\label{conclusion}

Vainshtein screening is crucial to pass the stringent tests of gravity available at local scales. We studied the Vainshtein regime in the most general subset of DHOST theories that evade all of the constraints from gravitational wave observations, i.e., for which gravitons propagate at the speed of light and do not decay. 

For non-zero values of the parameter that characterizes the higher-order derivatives in DHOST theories, i.e.~$\beta_1$, the screening mechanism is broken. Negative values of $\beta_1$ are ruled out because the gravitational potentials inside matter do not scale as the inverse of the distance. For positive values of $\beta_1$,  the scalar field can be neglected inside matter 
while its non-lineartities contribute to the potentials {outside of matter}. Therefore, we find that the gravitational potentials scale as the inverse of the distance but have different gravitational constants inside and outside of the body, and between themselves.

For generic values of the other parameters, in particular of $\alphaB$, $\alphaM$  and $\dot \beta_1$, the measurement of the Shapiro time-delay with the Cassini spacecraft constraints the parameter $\beta_1$ to be smaller than $10^{-5}$. But there is a special region of the parameter space where the two gravitational potentials are equal outside of the body and the bound from Cassini is evaded. Here the constraint on
$\beta_1$ weakens, $\beta_1< 7.5 \times 10^{-3}$ at $2\sigma$, and is obtained from the measurement of the orbital decay rate of the Hulse-Taylor pulsar.

{Theories conformally related to General Relativity have been considered separately. Generically, they are ruled out unless $\beta_1=0$. For specific values of the parameters $\alphaB$ and $\alphaM$ they admit $0 \le \beta_1 \lesssim 10^{-6}$.}

In App.~\ref{otherconstraints} we studied also other constraints, coming from the stellar structure, that can be imposed using the difference between the Newtonian potential inside and outside of a star. However, they all turned out to be weaker than the bounds obtained using the exterior solutions. These constraints can be improved with more sophisticated methods (see e.g.~\cite{Saltas:2018mxc}) and with more accurate data that will be available in the future.

\vskip.1cm

\emph{Note added.---}  Another article~\cite{Hirano:2019scf}, whose content overlaps with ours, appeared while finalizing this article.

%
%
%

\section*{Acknowledgements}

We thank the authors of \cite{Hirano:2019scf} for interesting correspondence.  MC thanks Francesco Villante, Santiago Casas, and Virginia Ajani for useful discussions.  MC is supported by the Labex P2IO and the Enhanced Eurotalents Fellowship.   ML acknowledges financial support from the Enhanced Eurotalents fellowship, a Marie Sklodowska-Curie Actions Programme, and the European Research Council under ERC-STG-639729, \emph{preQFT: Strategic Predictions for Quantum Field Theories}.

%
%
%
\appendix

\section{Coefficients and equations} 
\label{sec:coeff}
We report here the explicit expressions of the coefficients introduced in the text and of some of the long expressions for the field equations.

\subsection{Action and field equations} 
\label{coefficientsapp}
The coefficients appearing in the action expanded in perturbations, \eqn{Lagrangians}, are given explicitly by 
 \be
 \begin{split}
c_1 & = -4H \alphaB  + H ( 4 \alphaH - 2 \beta_3 (1 + \alphaM )) - 2 \dot \beta_3 \;,  \\
c_2 & = 4 H (1 + \alphaM - c_{\rm T}^2)  +  4 \left(  H    \alphaH (1 + \alphaM)   + \dot \alpha_{\rm H} \right)\;,  \\
c_3 & = - 2 H^2 \mathcal{C}_2+ \frac{1}{2} \left\{   H \left[ 4 \dot \alpha_{\rm H} - 2 ( 1 + \alphaM) \dot \beta_3  - \beta_3 \dot {\alpha}_{\rm M} \right] - \ddot \beta_3           \right\} \\ 
& \quad \quad + \frac{1}{2} \big\{ - H^2 ( 1 + \alphaM) \left[ - 4 \alphaH + \beta_3 ( 1 + \alphaM)\right]   \\
& \quad \quad  + 4 \alphaH \dot H - \beta_3 ( 1 + \alphaM) \dot H  \big\} \;, \\
 c_4 & = 4 ( 1 + \alpha_{\rm H} ) \ ,  \qquad  c_5  = -2 c_{\rm T}^2 \ , \qquad  c_6  = - \beta_3 \ , \\ 
 c_7  &= 4 \alpha_{\rm H} \;, \qquad
 c_8  = -2 (  2 \beta_1 + \beta_3 ) \ , \qquad  c_9  = 4 \beta_1 + \beta_3 \ , 
 \end{split}
 \ee
 with
\be
\mathcal{C}_2 \equiv  - \alphaM + \alphaB ( 1 + \alphaM ) + \alphaT   +  ( 1 + \alphaB)\frac{\dot H}{H^2} + \frac{ \dot \alpha_{\rm B} }{H} + \frac{\bar \rho_{\rm m}}{2H^2 M^2}   \ ,
\ee
 \be
 \begin{split}
 b_1 & =  H  \left[ 4 \alphaB + \alphaV (-1+\alphaM)  -2 \alphaM + 3 \alphaT \right] + \dot \alpha_{\rm V} \\
 & \quad \quad - H \left[  8 \beta_1 \alphaM + \alphaH ( 3 + \alphaM ) \right]  - \dot \alpha_{\rm H} - 8 \dot \beta_1 \;, \\
 b_2 & = \alpha_{\rm V} - \alpha_{\rm H} - 4 \beta_1 \ , \qquad  b_3  = c_{\rm T}^2 -1  \ ,  \\
 b_4  &= -c_7 \ ,  \qquad b_5  = -c_8 \ , \qquad  b_6  = -2 c_9 \ ,  
 \end{split}
\ee
and
\be
d_1  = -b_3 - b_2 \ , \qquad  d_2  = c_9 \;.
\ee

Let us define 
\begin{align}
\begin{split}
C_1 & \equiv  \frac{1}{4} \left( c_1  - H c_8 ( 1 + \alphaM)  - \dot c_8 \right) \;,\\
C_2 & \equiv \frac{1}{4} \left(   c_2 - H c_7 ( 1 + \alphaM ) - \dot c_7 \right)\;, \\
C_3 & \equiv  \frac{1}{4} \left\{ 2 c_3 + (1 + \alphaM) \left[ 2 H \dot c_9 + c_9 \left( H^2 (1 + \alphaM)  + \dot H \right)  \right] \right. \\
& \quad \quad \quad \left. + c_9 H \dot \alpha_{\rm M} + \ddot c_9     \right\}  \;, \\
C_4 & \equiv \half ( c_9 H ( 1 + \alphaM) + \dot c_9 ) \;.
\end{split}
\end{align}
The variation of the action \eqn{EFTactionexp} gives the field equations \eqn{fieldeq}, with
\begin{align} \label{phieq1}
\begin{split}
&{\cal E}_\Phi = C_1\partial^2 \pi - \frac{c_8}{4}  \partial^2 \dot \pi + \frac{c_6}{2} \partial^2 \Phi   + \frac{c_4}{4}  \partial^2 \Psi 
\\
& \hspace{.25in} +  \frac{1}{4} \left[  \frac{b_2}{a^2} Q_2 [\pi,\pi ] - \frac{b_5}{a^2} \partial_i \left( \partial_j \pi \partial_i \partial_j \pi \right) \right] -\frac{a^2 \delta \rho_{\rm m}}{2 M^2}\ , 
\end{split}
\end{align}
\begin{align} \label{psieq1}
\begin{split}
&  {\cal E}_\Psi = C_2 \partial^2 \pi - \frac{c_7}{4}  \partial^2 \dot \pi + \frac{c_4}{4}  \partial^2 \Phi  + \frac{c_5}{2}  \partial^2 \Psi \\
& \hspace{.8in} +  \frac{1}{4} \left[  \frac{b_3}{a^2} Q_2 [\pi,\pi ] - \frac{b_4}{a^2} \partial_i \left( \partial_j \pi \partial_i \partial_j \pi \right) \right] \ , 
\end{split}
\end{align}
and 
\begin{align}\label{pieq1}
\begin{split}
 &  {\cal E}_\pi = C_3  \partial^2 \pi +  C_4 \partial^2 \dot \pi + \frac{c_9}{2} \partial^2 \ddot \pi  +  \frac{c_1}{4}  \partial^2 \Phi     + \frac{c_8}{4}  \partial^2 \dot \Phi \\
&+  \frac{c_2}{4}  \partial^2 \Psi + \frac{c_7}{4}  \partial^2 \dot \Psi    +  \frac{1}{4 a^2}  Q_2 [ \pi , b_1\pi+ 2b_2 \Phi + 2b_3 \Psi ] \\
&  + \frac{1}{4 a^2} \partial_i \left[ \partial_i \pi  \, \partial^2 ( b_4 \Psi + b_5 \Phi  + b_6 \dot \pi) \right] \\
&   + \frac{1}{8 a M^2} \frac{d}{d t} \left( \frac{M^2 b_6 }{ a} \right) \partial^2 \left( \partial \pi \right)^2  + \frac{b_6}{4 a^2 } \partial^2 (  \partial_i \pi \partial_i \dot \pi  )  \\
&  + \frac{d_1}{4 a^4} Q_3[\pi , \pi , \pi]   + \frac{d_2}{4 a^4} \partial_i \left[ \partial_i \pi \partial^2 ( \partial \pi )^2 \right] \ ,
\end{split}
\end{align}
where we have defined $Q_3[ \varphi_a, \varphi_b  , \varphi_c ]  \equiv  \varepsilon^{ikm} \varepsilon^{jln} \partial_i \partial_j \varphi_a \partial_k \partial_l \varphi_b \partial_m \partial_n \varphi_c$.
Note that these equations are general: no degeneracy conditions or observational constraints have been assumed.

\subsection{{Linear solutions}}
\label{coefficientsapp2}
Here we give the coefficients relevant for the linear solutions in \eqn{linearsolutions}.  These can be written quite compactly in terms of the coefficients in the linear solution for $\pi$ \eqn{linearpi}, which we define as 
\be
\mu_\pi = - \frac{a^2}{2 \mpl} \frac{\nufive }{\nu_2}  \ , \quad \quad \nu_\pi = - \frac{a^2}{2 \mpl} \frac{\nusix }{\nu_2}  \ .
\ee
Then, in terms of these, the coefficients in \eqn{linearsolutions} are,
\begin{align}
\begin{split}
\mu_\Phi & = \frac{a^2}{2 M^2 (1-\beta_1)^2 } + \frac{   \mu_\pi \varpi_\Phi       - \dot \mu_\pi \beta_1     }{1-\beta_1}  \ ,  \\
\nu_\Phi & = \frac{  -\mu_\pi \beta_1  + \nu_\pi \varpi_\Phi      - \dot \nu_\pi \beta_1          }{1-\beta_1} \ , \quad \sigma_\Phi =  \frac{- \nu_ \pi \beta_1}{ 1 - \beta_1} \ ,   \\
\end{split}
\end{align}
and
\begin{align}
\begin{split}
\mu_\Psi & = \frac{a^2  (1 - 2 \beta_1) }{2 M^2 (1-\beta_1)^2 }  +  \frac{   \mu_\pi \varpi_\Psi  + \dot \mu_\pi \beta_1          }{1-\beta_1} \ , \\
\nu_\Psi & = \frac{    \mu_\pi \beta_1  + \nu_\pi \varpi_\Psi  + \dot \nu_\pi \beta_1      }{1-\beta_1}    \ ,  \quad \sigma_\Psi  =\frac{  \nu_\pi  \beta_1}{1-\beta_1} \ , 
\end{split}
\end{align}
where 
\begin{align}
\begin{split}
\varpi_\Phi & =  \frac{  H \left( \alphaB - \alphaM  + \beta_1 ( 1 + \alphaM)  \right)  - \dot \beta_1 }{1-\beta_1} \ , \\
\varpi_\Psi & = \frac{1}{1 - \beta_1} \big[ H\left( \alphaB + \beta_1 (1-2 \alphaB + \alphaM) - \beta_1^2 (2 + \alphaM) \right)  \\
& \qquad \qquad  - \dot \beta_1 (1 - 2 \beta_1) \big]  \ . 
\end{split}
\end{align}

%
%
%
\subsection{{Field equations in spherical symmetry}}
\label{coefficientsapp3}
In spherically symmetry, the field equations \eqn{phieq1}--\eqn{pieq1} reduce to
\begin{align} \label{potspher1}
\begin{split}
& 4 C_1 x - c_8(\dot x + 2 H x) + 2 c_6 y + c_4 z \\
& \hspace{.6in}  + \Lambda^3 \left[ 2 b_2  x^2 - b_5  x ( x + r x ') \right]= 4  \frac{\Mp \mathcal{ A}}{M^2}  \;, 
\end{split}
\end{align}
\begin{align} \label{potspher2}
\begin{split}
&4 C_2 x  - c_7  (\dot x + 2 H x)  + c_4 y + 2 c_5 z   \\
& \hspace{.9in} + \Lambda^3 \left[ 2 b_3 x^2 - b_4  x ( x + r x' ) \right] =0 \;  ,
\end{split}
\end{align}
and 
\begin{align}
\begin{split} \label{pispher}
& 4 C_3 x + 4 C_4 ( \dot x + 2 H x ) + 2 c_9 ( \ddot x + 4 H \dot x + 2 x ( 2 H^2 + \dot H ) )  \\
&  + c_1 y + c_8 ( \dot y + 2 H y ) + c_2 z + c_7 ( \dot z + 2 H z) + 2 b_1 \Lambda^3 x^2  \\  
&   + \Lambda^3 \Big[ 4  x ( b_2 y + b_3 z)  +  x \left( b_4 ( 3 z + r z' ) + b_5 ( 3 y +r y' )  \right) \\
&  +  x  b_6 ( 3 \dot x + 6 H x + 2 H r  x ' + r \dot x ')    +  \frac{a}{M^2} \frac{d }{d t} \left( \frac{ M^2 b_6}{ a} \right)   x ( x + r x') \\
&   + b_6  ( r \dot x x' + 4 H x ( x + r x') + x ( 2 \dot x + r \dot x ') )  \Big] \\
&  + 2 \Lambda^6 \left[ d_1  x^3 + d_2  x ( 3 x^2 + r^2 x'^2 + 6 r x x' + r^2 x x'' ) \right] =0 \;.
\end{split}
\end{align}

%

\section{Stellar constraints} \label{otherconstraints}

In this appendix, we use properties of stars to constrain the relative difference between the gravitational constant inside and outside of matter,
\be
\epsilonMatt_G \equiv \frac{G_\Phi^{\rm in} - G_\Phi^{\rm out}}{G_\Phi^{\rm out}} \approx \epsilon^{\rm in}_\Phi - \epsilon^{\rm out}_\Phi =  \frac{\beta_1\upsilon }{2 \xi (1-\beta_1)^2}     \ . 
\ee

We assume that masses of stars in the literature are measured using the value of the gravitational constant outside of a source, i.e.~$G_\Phi^{\rm out} = G_* (1+\epsilon_\Phi^{\rm out})$ in \eqn{yzsols}, and that this is the gravitational constant value measured on the Earth, $G_\Phi^{\rm out} = 6.6276 \times 10^{-8} \text{ dyne cm}^2/\text{gm}^2$.  

Here, we show that the constraints involving the interior of the star are weaker than the ones presented in \secref{constraintssec}.  
The first constraint is an upper bound and comes from the Chandrasekhar mass limit. The second comes from the minimum main-sequence mass of brown dwarfs and  gives a lower bound.  We also present a third constraint, coming from fitting the mass-radius relation for white dwarfs.  
This gives the best bound, i.e.,
\be
-0.060 \lesssim \epsilonMatt_G \lesssim 0.031 \, \text{  at  } \, \,  2\sigma.  
\ee
{Of course, with improved observational data or theoretical modeling, these bounds could be improved in the future.}

%
%
%
\subsection{Chandrasekhar mass limit}

First, we consider the Chandrasekhar mass limit, which is the largest mass that a white dwarf can have \cite{Chandrasekhar:1931ftj,Chandrasekhar:1935zz}.  This limit exists because if the white dwarf had a larger mass, the electron degeneracy pressure would not be able to support the star, and it would collapse into a neutron star or a black hole.  (Reference \cite{Jain:2015edg} considered a similar constraint in the context of the DHOST theories which have a modification of the Newtonian potential \eqn{leadinginside} proportional to $ m''$.) 

The Chandrasekhar mass limit $M_{\rm Ch}$ for white dwarfs is proportional to $(G^{\rm in}_\Phi)^{-3/2}$ \cite{Chandrasekhar:1935zz}, and the current theoretical value of this limit is $M_{\rm Ch} \approx 1.44 M_\odot$.  Taking into account the difference between the Newton constants inside and outside of the star, this limit becomes $M_{\rm Ch} \approx 1.44 M_\odot ( 1 + \epsilonMatt_G)^{-3/2}$.  The largest white dwarf that we have seen has a mass of $m_{\rm tot} = ( 1.37 \pm 0.01) M_\odot$.  Because the Chandrasekhar limit cannot be below the heaviest white dwarf that we have seen, we must have $m_{\rm tot} \lesssim M_{\rm Ch}$, which in our theory translates to
\be
\epsilonMatt_G \lesssim 0.039  \ . 
\ee

%
%

\subsection{Brown dwarfs} \label{browndwarfsec}
Next, we move on to consider constraints coming from the burning process in brown dwarfs and red dwarfs \cite{Burrows:1992fg}.  This bound is based on the fact that the luminosity generated in the interior of the star from hydrogen burning, $L_{\rm HB}$, must equal the luminosity emitted by the star from the photosphere, $L_e$.  This analysis gives a smallest mass, the minimum main-sequence mass $M_{\rm MS}$, that is consistent with having a stable burning process in the body of the star.  In this work, we use the results of \cite{Burrows:1992fg}, but keep track of the dependence on $G^{\rm in}_\Phi$: the interested reader can find many more details in \cite{Burrows:1992fg}.  (Reference \cite{Sakstein:2015zoa} has used a similar argument to constrain modification of the Newtonian potential proportional to $ m''$.)  

To find the luminosity in hydrogen burning, we start with the equation for hydrostatic equilibrium $d P / dr = - G^{\rm in }_\Phi m ( r ) \rho ( r ) / r^2$, where $P$ is the pressure, $\rho$ is the mass density, and $m(r)$ is defined in \eqn{mdef} (although we neglect the expansion of the universe on these small scales).  Then, we assume an equation of state $P = K( \eta) \rho^{5/3}$, where $\eta \equiv c_\eta \rho^{2/3} / T$ (for $c_\eta$ a constant) is a measure of the degeneracy of the electron gas and is constant throughout the star.  The numerical values of both $K(\eta)$ and $c_\eta$ can be determined from fundamental parameters such as $\hbar$, the electron mass, the hydrogen mass, and the number of baryons per electron. {This allows us to find an equation for the density profile, which we write as $\rho(r) = \rho_c \theta(r)^{3/2}$.  After also defining $\tau = r / r_*$ and $r_*^2 = 5 \rho_c^{-1/3} K(\eta) / ( 8 \pi G_\Phi^{\rm in})$, we obtain the Lane-Emden equation
\be \label{leeq}
\frac{d}{d \tau } \left( \tau^2 \frac{d \theta}{d \tau} \right) = - \tau^2 \theta^{3/2}
\ee
where at $\tau = 0$, we have both $\theta = 1$ and $d \theta / d \tau = 0$.  
}

For the total luminosity in hydrogen burning, we obtain
\begin{align}
\begin{split} \label{lumhb}
L_{\rm HB} &= 7.53 \times 10^4 L_\odot \, ( 1 + \epsilonMatt_G)^{16.466} \left( \frac{m_{\rm tot}}{0.1 M_\odot} \right)^{11.977} \\
& \quad \quad \times \eta^{-6.316} \left( 1 + \alpha/\eta \right)^{-16.466} \ ,
\end{split}
\end{align}
where $\alpha = 4.82$, and for the luminosity emitted from the surface, we obtain
\begin{align}
L_e & = 3.71 L_\odot  ( 1 + \epsilonMatt_G)^{1.549} \left( \frac{m_{\rm tot} }{0.1 M_\odot} \right)^{1.305} \left( \frac{ 10^{-2}}{\kappa_R} \right)^{1.184} \nonumber \\
&\quad \quad \times \eta^{-4.352} \left( 1 + \alpha / \eta \right)^{-0.36} \ ,   \label{lumemit}
\end{align}
where $\kappa_R \sim 10^{-2}$ (in units of $\text{ cm}^2/\text{gm}$) is the Rosseland mean opacity, which determines the optical depth of the star.

The condition to have stable burning is that $L_{\rm HB} = L_e$, which means that,
\be
(1+\epsilonMatt_G)^{1.398} \frac{m_{\rm tot}}{0.1 M_\odot} \left( \frac{ \kappa_R}{10^{-2}} \right)^{0.111}  = 0.3948 \, I( \eta) \ , 
\ee
where $I( \eta ) \equiv \eta^{0.184} ( 1 + \alpha / \eta )^{1.509}$.  As discussed in \cite{Burrows:1992fg}, the function $I(\eta)$ has a minimum value of 2.337 at $\eta = 34.7$, so that we have the bound 
\be
(1+\epsilonMatt_G)^{1.398} \frac{m_{\rm tot}}{0.1 M_\odot} \left( \frac{ \kappa_R}{10^{-2}} \right)^{0.111}  \gtrsim 0.9227  \ .
\ee
Now, the smallest red dwarf that has been measured has a mass of $m_{\rm tot} = ( 0.093 \pm 0.0008) M_\odot$ \cite{Segransan:2000jq}. 
 Thus, we finally have 
\be
 \epsilonMatt_G \gtrsim -0.0057 + 0.9943 \bigg[  \left( \frac{10^{-2}}{\kappa_R} \right)^{0.0799} - 1 \bigg]  \ .
\ee
Notice that the bound depends on the mean opacity $\kappa_R$.  Previous results in the literature \cite{Burrows:1992fg,Sakstein:2015zoa} have set $\kappa_R = 10^{-2} \text{ cm}^2/\text{gm}$ and noted the relatively weak dependence as a justification.  However, even though the dependence is quite weak, it can have a fairly large impact on the final bound.  For example, in \cite{Freedman:2007cm}, we see that for $\rho_e \sim 10^{-5}-10^{-4} \text{ gm}/\text{cm}^2$ and $T_e \sim 2000 \text{ K}$, that $10^{-2 } \lesssim \kappa_R \lesssim 10^{-1}$. Thus, we obtain a range of bounds, 
\begin{align}
\begin{split}
& \epsilonMatt_G \gtrsim -0.0057 \ , \, \, \,  \text{for} \,  \, \, \kappa_R = 10^{-2} \ ,  \\
& \epsilonMatt_G \gtrsim -0.173  \ , \, \, \,  \text{for} \,  \, \, \kappa_R = 10^{-1} \ .
\end{split}
\end{align}

\begin{figure}
\centering
\includegraphics[width=8.6cm]{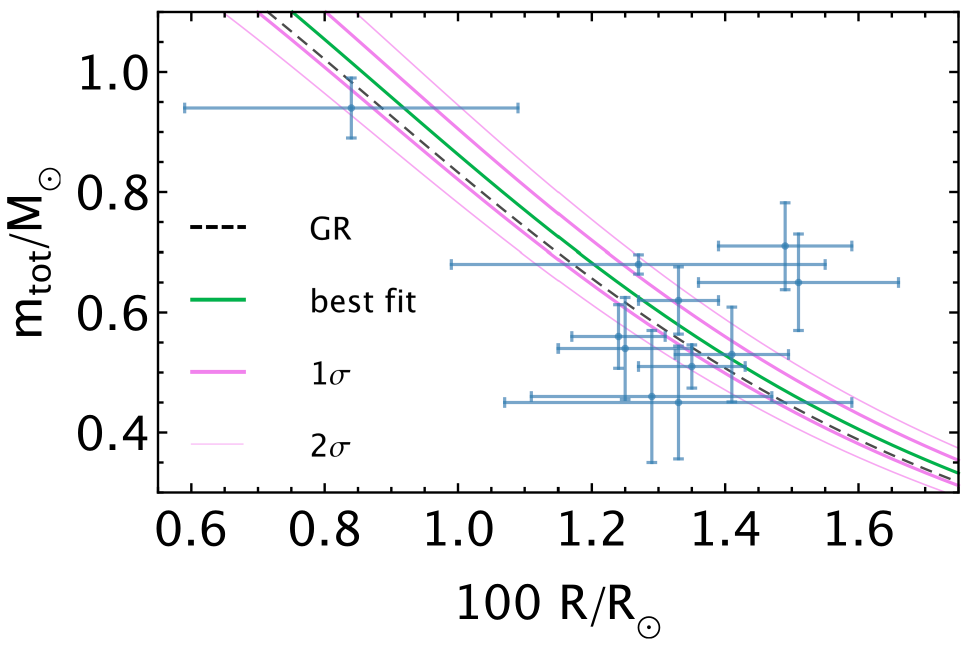}
\caption{ In this figure, we compare our predictions for the mass-radius relation for white dwarfs with data from \cite{Holberg:2012pu} (which is shown above as the blue data points and error bars).  At $1\sigma$, we find $-0.039 \lesssim \epsilonMatt_G \lesssim 0.0066$, and at $2\sigma$ we find $-0.060 \lesssim \epsilonMatt_G \lesssim 0.031$.    }
\label{fig:1}
\end{figure}

%
%
%
\subsection{White dwarfs} \label{wdsec}

Finally, we move on to bounds set by the mass-radius relation of white dwarfs by comparing to the catalogue of stars in \cite{Holberg:2012pu}.  To simplify our analysis, we assume that the stars are each made of a low temperature, completely degenerate Fermi gas, and follow \cite{shapiro_teukolsky_1983}.  Practically speaking, this assumption means that the profile of the star is not significantly affected by the temperature, an assumption which is reasonable for $T \lesssim 30,000 \text{ K}$ \cite{Holberg:2012pu}.  For this reason, we omit the data point with $T = 49,000 \text{ K}$ in \cite{Holberg:2012pu}.  See \cite{Saltas:2018mxc} for an example of a more involved analysis which includes these non-trivial temperature effects.

Under these assumptions, the equation of state of the star is given by,
\begin{align}
P(s) &= P_0 \phi ( s ) \ ,  \quad \rho ( s )  = \rho_0 s^3   \ ,\label{pofx}\\
\phi(s) & = \frac{1}{8 \pi^2} \left[ s \sqrt{ 1 + s^2 } \left( \frac{2 s^2}{3} -1\right) + \log \left( s + \sqrt{ 1 + s^2} \right) \right] \ ,  \nonumber
\end{align}
where $P_0 =  1.4218 \times 10^{25} \text{dyne}\, \, \text{cm}^{-2}$, $\rho_0 = 1.9479 \times 10^6 \text{gm} \, \, \text{cm}^{-3}$, and $s \equiv p_{\rm F} /(m_e c)$ is the unitless Fermi momentum of an electron, which is in general a function of the distance $r$ from the center of the star, i.e. $s = s(r)$.  The equation for hydrostatic equilibrium is the same as that mentioned above \eqn{lumhb}, and combined with the mass conservation equation, $m' ( r ) = 4 \pi r^2 \rho ( r )$, one obtains a system of first order differential equations for $s(r)$ and $m(r)$.

After defining $ q \equiv r / R_\odot$, $\vartheta \equiv  m/ m_0$, $m_0 \equiv (4 \pi R_\odot^3 \rho_0 / 3)$, $\gamma \equiv P_0 / (4 \pi G_\Phi^{\rm in} \rho_0^2 R_\odot^2 / 3 )$ and $y(q) \equiv s(q)^2$, this system can be written as
 \begin{align}
 \begin{split}
 \vartheta'(q) &  = 3 q^2 y(q)^{3/2} \;, \\
 y'(q) & = 6 \pi^2 \gamma^{-1} ( 1 + \epsilonMatt_G ) \frac{\vartheta( q) \sqrt{1+ y( q) }   }{q^2}  \ .
 \end{split}
 \end{align}
The initial conditions are $\vartheta( 0 ) = 0$ and $y(0 ) = y_0$ (which determines the total mass of the star), and the radius of the star $R$ is given by $y(R) = 0$.  In particular, we have $\gamma = 0.00277$, and $M_\odot / m_0 = 7.24 \times 10^{-7}$.

To do the fit to the data, we first find the points $\tilde R_i$ on the theoretical curve that match best with each data point $(R_i , M_i)$.  In particular, we define 
\be
\chi_i^2 ( R ; \epsilonMatt_G )  = \frac{\left( m_{\rm tot} ( R; \epsilonMatt_G ) -  M_i \right)^2}{\sigma_{M,i}^2} + \frac{ \left( R - R_i \right) }{\sigma_{R,i}^2} \;,
\ee
where $M_i$, $\sigma_{M,i}$, $R_i$, and $\sigma_{R,i}$ are respectively the mass, error bar for the mass, radius, and error bar for the radius of the $i$th star, and we minimize each $\chi^2_i$ to find the $\tilde R_i$ where it is minimum.  The total $\chi^2$ is then given by 
\be
\chi^2 ( \epsilonMatt_G ) = \sum_{i=1}^N \chi^2_i ( \tilde R_i ; \epsilonMatt_G) \label{chisq} \ ,
\ee
where for us $N = 12$.  We then minimize \eqn{chisq} with respect to $\epsilonMatt_G$ to find our constraints.  

Calling $\tilde \epsilonMatt_G$ the minimum, we then determine the $n\sigma$ region by finding where $\Delta \chi^2 ( \epsilonMatt_G) \equiv \chi^2 (\epsilonMatt_G) - \chi^2 ( \tilde \epsilonMatt_G) = n^2$.  In particular, we find a best fit value of $\epsilonMatt_G = - 0.017$, a $1\sigma$ range of $-0.039 \lesssim \epsilonMatt_G \lesssim 0.0066$, and a $2\sigma$ range of $-0.060 \lesssim \epsilonMatt_G \lesssim 0.031$.   We present our results in \figref{fig:1}.

%
%
%

\newpage

 \bibliographystyle{utphys}
\bibliography{EFT_DE_biblio4}

\providecommand{\href}[2]{#2}\begingroup\raggedright\begin{thebibliography}{10}

\bibitem{Horndeski:1974wa}
G.~W. Horndeski, ``{Second-order scalar-tensor field equations in a
  four-dimensional space},'' {\em Int.J.Theor.Phys.} {\bf 10} (1974) 363--384.

\bibitem{Deffayet:2011gz}
C.~Deffayet, X.~Gao, D.~Steer, and G.~Zahariade, ``{From k-essence to
  generalised Galileons},'' {\em Phys.Rev.} {\bf D84} (2011) 064039,
  \href{http://xxx.lanl.gov/abs/1103.3260}{{\tt 1103.3260}}.

\bibitem{Zumalacarregui:2013pma}
M.~Zumalac{\'a}rregui and J.~Garc{\'\i}a-Bellido, ``{Transforming gravity: from
  derivative couplings to matter to second-order scalar-tensor theories beyond
  the Horndeski Lagrangian},'' {\em Phys.Rev.} {\bf D89} (2014), no.~6 064046,
  \href{http://xxx.lanl.gov/abs/1308.4685}{{\tt 1308.4685}}.

\bibitem{Gleyzes:2014dya}
J.~Gleyzes, D.~Langlois, F.~Piazza, and F.~Vernizzi, ``{Healthy theories beyond
  Horndeski},'' {\em Phys. Rev. Lett.} {\bf 114} (2015), no.~21 211101,
  \href{http://xxx.lanl.gov/abs/1404.6495}{{\tt 1404.6495}}.

\bibitem{Langlois:2015cwa}
D.~Langlois and K.~Noui, ``{Degenerate higher derivative theories beyond
  Horndeski: evading the Ostrogradski instability},'' {\em JCAP} {\bf 1602}
  (2016), no.~02 034, \href{http://xxx.lanl.gov/abs/1510.06930}{{\tt
  1510.06930}}.

\bibitem{Crisostomi:2016czh}
M.~Crisostomi, K.~Koyama, and G.~Tasinato, ``{Extended Scalar-Tensor Theories
  of Gravity},'' {\em JCAP} {\bf 1604} (2016), no.~04 044,
  \href{http://xxx.lanl.gov/abs/1602.03119}{{\tt 1602.03119}}.

\bibitem{BenAchour:2016fzp}
J.~Ben~Achour, M.~Crisostomi, K.~Koyama, D.~Langlois, K.~Noui, and G.~Tasinato,
  ``{Degenerate higher order scalar-tensor theories beyond Horndeski up to
  cubic order},'' {\em JHEP} {\bf 12} (2016) 100,
  \href{http://xxx.lanl.gov/abs/1608.08135}{{\tt 1608.08135}}.

\bibitem{Langlois:2018dxi}
D.~Langlois, ``{Dark Energy and Modified Gravity in Degenerate Higher-Order
  Scalar-Tensor (DHOST) theories: a review},''
  \href{http://xxx.lanl.gov/abs/1811.06271}{{\tt 1811.06271}}.

\bibitem{Kobayashi:2019hrl}
T.~Kobayashi, ``{Horndeski theory and beyond: a review},''
  \href{http://xxx.lanl.gov/abs/1901.07183}{{\tt 1901.07183}}.

\bibitem{Will:2014xja}
C.~M. Will, ``{The Confrontation between General Relativity and Experiment},''
  {\em Living Rev.Rel.} {\bf 17} (2014) 4,
  \href{http://xxx.lanl.gov/abs/1403.7377}{{\tt 1403.7377}}.

\bibitem{Vainshtein:1972sx}
A.~I. Vainshtein, ``{To the problem of nonvanishing gravitation mass},'' {\em
  Phys. Lett.} {\bf 39B} (1972) 393--394.

\bibitem{Babichev:2013usa}
E.~Babichev and C.~Deffayet, ``{An introduction to the Vainshtein mechanism},''
  {\em Class. Quant. Grav.} {\bf 30} (2013) 184001,
  \href{http://xxx.lanl.gov/abs/1304.7240}{{\tt 1304.7240}}.

\bibitem{TheLIGOScientific:2017qsa}
{\bf Virgo, LIGO Scientific} Collaboration, B.~P. Abbott {\em et.~al.},
  ``{GW170817: Observation of Gravitational Waves from a Binary Neutron Star
  Inspiral},'' {\em Phys. Rev. Lett.} {\bf 119} (2017), no.~16 161101,
  \href{http://xxx.lanl.gov/abs/1710.05832}{{\tt 1710.05832}}.

\bibitem{Creminelli:2017sry}
P.~Creminelli and F.~Vernizzi, ``{Dark Energy after GW170817 and GRB170817A},''
  {\em Phys. Rev. Lett.} {\bf 119} (2017), no.~25 251302,
  \href{http://xxx.lanl.gov/abs/1710.05877}{{\tt 1710.05877}}.

\bibitem{Sakstein:2017xjx}
J.~Sakstein and B.~Jain, ``{Implications of the Neutron Star Merger GW170817
  for Cosmological Scalar-Tensor Theories},'' {\em Phys. Rev. Lett.} {\bf 119}
  (2017), no.~25 251303, \href{http://xxx.lanl.gov/abs/1710.05893}{{\tt
  1710.05893}}.

\bibitem{Ezquiaga:2017ekz}
J.~M. Ezquiaga and M.~Zumalacrregui, ``{Dark Energy After GW170817: Dead Ends
  and the Road Ahead},'' {\em Phys. Rev. Lett.} {\bf 119} (2017), no.~25
  251304, \href{http://xxx.lanl.gov/abs/1710.05901}{{\tt 1710.05901}}.

\bibitem{Baker:2017hug}
T.~Baker, E.~Bellini, P.~G. Ferreira, M.~Lagos, J.~Noller, and I.~Sawicki,
  ``{Strong constraints on cosmological gravity from GW170817 and GRB
  170817A},'' {\em Phys. Rev. Lett.} {\bf 119} (2017), no.~25 251301,
  \href{http://xxx.lanl.gov/abs/1710.06394}{{\tt 1710.06394}}.

\bibitem{deRham:2018red}
C.~de~Rham and S.~Melville, ``{Gravitational Rainbows: LIGO and Dark Energy at
  its Cutoff},'' {\em Phys. Rev. Lett.} {\bf 121} (2018), no.~22 221101,
  \href{http://xxx.lanl.gov/abs/1806.09417}{{\tt 1806.09417}}.

\bibitem{Gleyzes:2014qga}
J.~Gleyzes, D.~Langlois, F.~Piazza, and F.~Vernizzi, ``{Exploring gravitational
  theories beyond Horndeski},'' {\em JCAP} {\bf 1502} (2015) 018,
  \href{http://xxx.lanl.gov/abs/1408.1952}{{\tt 1408.1952}}.

\bibitem{Creminelli:2018xsv}
P.~Creminelli, M.~Lewandowski, G.~Tambalo, and F.~Vernizzi, ``{Gravitational
  Wave Decay into Dark Energy},'' {\em JCAP} {\bf 1812} (2018), no.~12 025,
  \href{http://xxx.lanl.gov/abs/1809.03484}{{\tt 1809.03484}}.

\bibitem{Bellazzini:2019xts}
B.~Bellazzini, M.~Lewandowski, and J.~Serra, ``{Amplitudes' Positivity, Weak
  Gravity Conjecture, and Modified Gravity},''
  \href{http://xxx.lanl.gov/abs/1902.03250}{{\tt 1902.03250}}.

\bibitem{Nicolis:2008in}
A.~Nicolis, R.~Rattazzi, and E.~Trincherini, ``{The Galileon as a local
  modification of gravity},'' {\em Phys. Rev.} {\bf D79} (2009) 064036,
  \href{http://xxx.lanl.gov/abs/0811.2197}{{\tt 0811.2197}}.

\bibitem{Kobayashi:2014ida}
T.~Kobayashi, Y.~Watanabe, and D.~Yamauchi, ``{Breaking of Vainshtein screening
  in scalar-tensor theories beyond Horndeski},'' {\em Phys. Rev.} {\bf D91}
  (2015), no.~6 064013, \href{http://xxx.lanl.gov/abs/1411.4130}{{\tt
  1411.4130}}.

\bibitem{Crisostomi:2017lbg}
M.~Crisostomi and K.~Koyama, ``{Vainshtein mechanism after GW170817},'' {\em
  Phys. Rev.} {\bf D97} (2018), no.~2 021301,
  \href{http://xxx.lanl.gov/abs/1711.06661}{{\tt 1711.06661}}.

\bibitem{Langlois:2017dyl}
D.~Langlois, R.~Saito, D.~Yamauchi, and K.~Noui, ``{Scalar-tensor theories and
  modified gravity in the wake of GW170817},'' {\em Phys. Rev.} {\bf D97}
  (2018), no.~6 061501, \href{http://xxx.lanl.gov/abs/1711.07403}{{\tt
  1711.07403}}.

\bibitem{Dima:2017pwp}
A.~Dima and F.~Vernizzi, ``{Vainshtein Screening in Scalar-Tensor Theories
  before and after GW170817: Constraints on Theories beyond Horndeski},''
  \href{http://xxx.lanl.gov/abs/1712.04731}{{\tt 1712.04731}}.

\bibitem{Bartolo:2017ibw}
N.~Bartolo, P.~Karmakar, S.~Matarrese, and M.~Scomparin, ``{Cosmic structures
  and gravitational waves in ghost-free scalar-tensor theories of gravity},''
  {\em JCAP} {\bf 1805} (2018), no.~05 048,
  \href{http://xxx.lanl.gov/abs/1712.04002}{{\tt 1712.04002}}.

\bibitem{Ganz:2018vzg}
A.~Ganz, N.~Bartolo, P.~Karmakar, and S.~Matarrese, ``{Gravity in mimetic
  scalar-tensor theories after GW170817},'' {\em JCAP} {\bf 1901} (2019),
  no.~01 056, \href{http://xxx.lanl.gov/abs/1809.03496}{{\tt 1809.03496}}.

\bibitem{Babichev:2018rfj}
E.~Babichev and A.~Lehbel, ``{The sound of DHOST},'' {\em JCAP} {\bf 1812}
  (2018), no.~12 027, \href{http://xxx.lanl.gov/abs/1810.09997}{{\tt
  1810.09997}}.

\bibitem{Kase:2018iwp}
R.~Kase and S.~Tsujikawa, ``{Dark energy scenario consistent with GW170817 in
  theories beyond Horndeski gravity},'' {\em Phys. Rev.} {\bf D97} (2018),
  no.~10 103501, \href{http://xxx.lanl.gov/abs/1802.02728}{{\tt 1802.02728}}.

\bibitem{Crisostomi:2017pjs}
M.~Crisostomi and K.~Koyama, ``{Self-accelerating universe in scalar-tensor
  theories after GW170817},'' {\em Phys. Rev.} {\bf D97} (2018), no.~8 084004,
  \href{http://xxx.lanl.gov/abs/1712.06556}{{\tt 1712.06556}}.

\bibitem{Crisostomi:2018bsp}
M.~Crisostomi, K.~Koyama, D.~Langlois, K.~Noui, and D.~A. Steer,
  ``{Cosmological evolution in DHOST theories},'' {\em JCAP} {\bf 1901} (2019),
  no.~01 030, \href{http://xxx.lanl.gov/abs/1810.12070}{{\tt 1810.12070}}.

\bibitem{Frusciante:2018tvu}
N.~Frusciante, R.~Kase, K.~Koyama, S.~Tsujikawa, and D.~Vernieri, ``{Tracker
  and scaling solutions in DHOST theories},'' {\em Phys. Lett.} {\bf B790}
  (2019) 167--175, \href{http://xxx.lanl.gov/abs/1812.05204}{{\tt 1812.05204}}.

\bibitem{Kase:2018aps}
R.~Kase and S.~Tsujikawa, ``{Dark energy in Horndeski theories after GW170817:
  A review},'' \href{http://xxx.lanl.gov/abs/1809.08735}{{\tt 1809.08735}}.

\bibitem{Ezquiaga:2018btd}
J.~M. Ezquiaga and M.~Zumalacrregui, ``{Dark Energy in light of Multi-Messenger
  Gravitational-Wave astronomy},'' {\em Front. Astron. Space Sci.} {\bf 5}
  (2018) 44, \href{http://xxx.lanl.gov/abs/1807.09241}{{\tt 1807.09241}}.

\bibitem{deRham:2016wji}
C.~de~Rham and A.~Matas, ``{Ostrogradsky in Theories with Multiple Fields},''
  {\em JCAP} {\bf 1606} (2016), no.~06 041,
  \href{http://xxx.lanl.gov/abs/1604.08638}{{\tt 1604.08638}}.

\bibitem{Langlois:2017mxy}
D.~Langlois, M.~Mancarella, K.~Noui, and F.~Vernizzi, ``{Effective Description
  of Higher-Order Scalar-Tensor Theories},'' {\em JCAP} {\bf 1705} (2017),
  no.~05 033, \href{http://xxx.lanl.gov/abs/1703.03797}{{\tt 1703.03797}}.

\bibitem{Bellini:2014fua}
E.~Bellini and I.~Sawicki, ``{Maximal freedom at minimum cost: linear
  large-scale structure in general modifications of gravity},'' {\em JCAP} {\bf
  1407} (2014) 050, \href{http://xxx.lanl.gov/abs/1404.3713}{{\tt 1404.3713}}.

\bibitem{Creminelli:2008wc}
P.~Creminelli, G.~D'Amico, J.~Norena, and F.~Vernizzi, ``{The Effective Theory
  of Quintessence: the $w<-1$ Side Unveiled},'' {\em JCAP} {\bf 0902} (2009)
  018, \href{http://xxx.lanl.gov/abs/0811.0827}{{\tt 0811.0827}}.

\bibitem{DAmico:2016ntq}
G.~D'Amico, Z.~Huang, M.~Mancarella, and F.~Vernizzi, ``{Weakening Gravity on
  Redshift-Survey Scales with Kinetic Matter Mixing},'' {\em JCAP} {\bf 1702}
  (2017) 014, \href{http://xxx.lanl.gov/abs/1609.01272}{{\tt 1609.01272}}.

\bibitem{Cusin:2017mzw}
G.~Cusin, M.~Lewandowski, and F.~Vernizzi, ``{Nonlinear Effective Theory of
  Dark Energy},'' {\em JCAP} {\bf 1804} (2018), no.~04 061,
  \href{http://xxx.lanl.gov/abs/1712.02782}{{\tt 1712.02782}}.

\bibitem{Hirano:2019nkz}
S.~Hirano, T.~Kobayashi, D.~Yamauchi, and S.~Yokoyama, ``{Constraining DHOST
  theories with linear growth of matter density fluctuations},''
  \href{http://xxx.lanl.gov/abs/1902.02946}{{\tt 1902.02946}}.

\bibitem{Kobayashi:2018xvr}
T.~Kobayashi and T.~Hiramatsu, ``{Relativistic stars in degenerate higher-order
  scalar-tensor theories after GW170817},'' {\em Phys. Rev.} {\bf D97} (2018),
  no.~10 104012, \href{http://xxx.lanl.gov/abs/1803.10510}{{\tt 1803.10510}}.

\bibitem{Achour:2016rkg}
J.~Ben~Achour, D.~Langlois, and K.~Noui, ``{Degenerate higher order
  scalar-tensor theories beyond Horndeski and disformal transformations},''
  {\em Phys. Rev.} {\bf D93} (2016), no.~12 124005,
  \href{http://xxx.lanl.gov/abs/1602.08398}{{\tt 1602.08398}}.

\bibitem{Brans:1961sx}
C.~Brans and R.~Dicke, ``{Mach's principle and a relativistic theory of
  gravitation},'' {\em Phys.Rev.} {\bf 124} (1961) 925--935.

\bibitem{Carroll:2004de}
S.~M. Carroll, A.~De~Felice, V.~Duvvuri, D.~A. Easson, M.~Trodden, and M.~S.
  Turner, ``{The Cosmology of generalized modified gravity models},'' {\em
  Phys. Rev.} {\bf D71} (2005) 063513,
  \href{http://xxx.lanl.gov/abs/astro-ph/0410031}{{\tt astro-ph/0410031}}.

\bibitem{Gleyzes:2015pma}
J.~Gleyzes, D.~Langlois, M.~Mancarella, and F.~Vernizzi, ``{Effective Theory of
  Interacting Dark Energy},'' {\em JCAP} {\bf 1508} (2015), no.~08 054,
  \href{http://xxx.lanl.gov/abs/1504.05481}{{\tt 1504.05481}}.

\bibitem{Jimenez:2015bwa}
J.~Beltran~Jimenez, F.~Piazza, and H.~Velten, ``{Evading the Vainshtein
  Mechanism with Anomalous Gravitational Wave Speed: Constraints on Modified
  Gravity from Binary Pulsars},'' {\em Phys. Rev. Lett.} {\bf 116} (2016),
  no.~6 061101, \href{http://xxx.lanl.gov/abs/1507.05047}{{\tt 1507.05047}}.

\bibitem{Weisberg:2010zz}
J.~M. Weisberg, D.~J. Nice, and J.~H. Taylor, ``{Timing Measurements of the
  Relativistic Binary Pulsar PSR B1913+16},'' {\em Astrophys. J.} {\bf 722}
  (2010) 1030--1034, \href{http://xxx.lanl.gov/abs/1011.0718}{{\tt 1011.0718}}.

\bibitem{Hulse:1974eb}
R.~A. Hulse and J.~H. Taylor, ``{Discovery of a pulsar in a binary system},''
  {\em Astrophys. J.} {\bf 195} (1975) L51--L53.

\bibitem{Bertotti:2003rm}
B.~Bertotti, L.~Iess, and P.~Tortora, ``{A test of general relativity using
  radio links with the Cassini spacecraft},'' {\em Nature} {\bf 425} (2003)
  374--376.

\bibitem{Saltas:2018mxc}
I.~D. Saltas, I.~Sawicki, and I.~Lopes, ``{White dwarfs and revelations},''
  {\em JCAP} {\bf 1805} (2018), no.~05 028,
  \href{http://xxx.lanl.gov/abs/1803.00541}{{\tt 1803.00541}}.

\bibitem{Hirano:2019scf}
S.~Hirano, T.~Kobayashi, and D.~Yamauchi, ``{On the screening mechanism in
  DHOST theories evading gravitational wave constraints},''
  \href{http://xxx.lanl.gov/abs/1903.08399}{{\tt 1903.08399}}.

\bibitem{Chandrasekhar:1931ftj}
S.~Chandrasekhar and E.~A. Milne, ``{The Highly Collapsed Configurations of a
  Stellar Mass},'' {\em Mon. Not. Roy. Astron. Soc.} {\bf 91} (1931), no.~5
  456--466.

\bibitem{Chandrasekhar:1935zz}
S.~Chandrasekhar, ``{The highly collapsed configurations of a stellar mass
  (Second paper)},'' {\em Mon. Not. Roy. Astron. Soc.} {\bf 95} (1935)
  207--225.

\bibitem{Jain:2015edg}
R.~K. Jain, C.~Kouvaris, and N.~G. Nielsen, ``{White Dwarf Critical Tests for
  Modified Gravity},'' {\em Phys. Rev. Lett.} {\bf 116} (2016), no.~15 151103,
  \href{http://xxx.lanl.gov/abs/1512.05946}{{\tt 1512.05946}}.

\bibitem{Burrows:1992fg}
A.~Burrows and J.~Liebert, ``{The Science of brown dwarfs},'' {\em Rev. Mod.
  Phys.} {\bf 65} (1993) 301--336.

\bibitem{Sakstein:2015zoa}
J.~Sakstein, ``{Hydrogen Burning in Low Mass Stars Constrains Scalar-Tensor
  Theories of Gravity},'' {\em Phys. Rev. Lett.} {\bf 115} (2015) 201101,
  \href{http://xxx.lanl.gov/abs/1510.05964}{{\tt 1510.05964}}.

\bibitem{Segransan:2000jq}
D.~Segransan, X.~Delfosse, T.~Forveille, J.~L. Beuzit, S.~Udry, C.~Perrier, and
  M.~Mayor, ``{Accurate masses of very low mass stars. 3. 16 New or improved
  masses},'' {\em Astron. Astrophys.} {\bf 364} (2000) 665,
  \href{http://xxx.lanl.gov/abs/astro-ph/0010585}{{\tt astro-ph/0010585}}.

\bibitem{Freedman:2007cm}
R.~S. Freedman, M.~S. Marley, and K.~Lodders, ``{Line and Mean Opacities for
  Ultracool Dwarfs and Extrasolar Planets},'' {\em Astrophys. J. Suppl.} {\bf
  174} (2008) 504, \href{http://xxx.lanl.gov/abs/0706.2374}{{\tt 0706.2374}}.

\bibitem{Holberg:2012pu}
J.~B. Holberg, T.~D. Oswalt, and M.~A. Barstow, ``{Observational Constraints on
  the Degenerate Mass-Radius Relation},'' {\em Astron. J.} {\bf 143} (2012) 68,
  \href{http://xxx.lanl.gov/abs/1201.3822}{{\tt 1201.3822}}.

\bibitem{shapiro_teukolsky_1983}
S.~A. Shapiro, Stuart L.;~Teukolsky, {\em Black Holes, White Dwarfs, and
  Neutron Stars}.
\newblock Wiley-Interscience, 1983.

\end{thebibliography}\endgroup

\end{document}